\documentclass[aps,prx,reprint,superscriptaddress]{revtex4-2}

\usepackage{dcolumn}
\usepackage{bm}

\usepackage{graphicx}
\usepackage{indentfirst}
\usepackage{braket}
\usepackage{float}
\usepackage{amsmath}
\usepackage{physics}
\usepackage{amssymb}
\usepackage{CJK}
\usepackage{esint}
\usepackage{color}
\usepackage[T1]{fontenc}
\usepackage{subfigure}
\usepackage{amsfonts}
\usepackage{footmisc}
\usepackage{scrextend}
\usepackage{multirow}
\usepackage[outdir=./]{epstopdf}
\usepackage{svg}
\usepackage{algorithm}
\usepackage{algpseudocode}
\usepackage{soul}
\usepackage{mathtools}
\usepackage[english]{babel}
\usepackage{url}
\usepackage{comment}
\usepackage{array}
\usepackage{subfiles}
\usepackage{tikz}
\usetikzlibrary{shapes,arrows}
\usepackage{mathrsfs}
\usepackage[mathscr]{eucal}
\usepackage[hyperfootnotes=false]{hyperref}
\usepackage{cleveref}
\usepackage{stmaryrd}

\definecolor{darkblue}{rgb}{0,0,0.5}
\hypersetup{
colorlinks=true,
linkcolor=black,
filecolor=blue,
citecolor=darkblue,  
urlcolor=black,
}

\usepackage{trimclip}

\makeatletter
\DeclareRobustCommand{\shortto}{%
  \mathrel{\mathpalette\short@to\relax}%
}

\newcommand{\short@to}[2]{%
  \mkern2mu
  \clipbox{{.5\width} 0 0 0}{$\m@th#1\vphantom{+}{\shortrightarrow}$}%
  }
\makeatother

\usepackage{pict2e,picture,graphicx}

\makeatletter
\DeclareRobustCommand{\Arrow}[1][]{%
\check@mathfonts
\if\relax\detokenize{#1}\relax
\settowidth{\dimen@}{$\m@th\rightarrow$}%
\else
\setlength{\dimen@}{#1}%
\fi
\sbox\z@{\usefont{U}{lasy}{m}{n}\symbol{41}}%
\begin{picture}(\dimen@,\ht\z@)
\roundcap
\put(\dimexpr\dimen@-.7\wd\z@,0){\usebox\z@}
\put(0,\fontdimen22\textfont2){\line(1,0){\dimen@}}
\end{picture}%
}
\makeatother

\def\be{\begin{equation}}
\def\ee{\end{equation}}
\def\ba{\begin{eqnarray}}
\def\ea{\end{eqnarray}}
\def\bal{\begin{equation}\begin{aligned}}
\def\eal{\end{aligned}\end{equation}}

\def\bp{\begin{pmatrix}}
\def\ep{\end{pmatrix}}

\usepackage{bm}

\newcommand{\calA}{{\cal A}}

\newcommand{\calC}{{\cal C}}

\newcommand{\calF}{{\cal F}}

\newcommand{\calU}{{\cal U}}

\usepackage[normalem]{ulem}

\begin{document}


\title{\textbf{Deterministic and Universal Frequency-Bin Gate for High-Dimensional Quantum Technologies} 
}%

\author{Xin Chen}%
 \email{chenxin@qzc.edu.cn}
\affiliation{%
 The College of Electrical and Information Engineering, Quzhou University, Quzhou 324000, China.
}%

\date{\today}




\begin{abstract}
High-dimensional photonic systems access large Hilbert spaces for quantum information processing. They offer proven advantages in quantum computation, communication, and sensing. However, implementing scalable, low-loss unitary gates across many
modes remains a central challenge. Here we propose a deterministic, universal,
and fully programmable high-dimensional quantum gate based on a
cavity-assisted sum-frequency–generation process, achieving near-unity
fidelity. The device implements an $M\times N$ truncated unitary transformation
($1\le M< N$), or a full unitary when $M=N$, on frequency-bin modes. With
current technology, the attainable dimensionality reaches
$M\times N\sim10^{4}$, with $N$ up to $10^{3}$, and can be further increased
using multiple pulse shapers. Combined with compatible SPDC sources,
high-efficiency detection, and fast feed-forward, this approach provides a
scalable, fiber-compatible platform for high-dimensional frequency-bin quantum
processing.
\end{abstract}


\maketitle

\section{Introduction}
Quantum information science has traditionally relied on \textit{qubits}---two-level systems that form the foundation of most existing architectures. Recent advances, however, highlight the benefits of accessing larger Hilbert spaces, either by scaling up the number of qubits or by employing multilevel (\(d\)-level) quantum systems (\textit{qudits}) with \(d>2\) \cite{Luo2014,wang2020qudits,lu2020quantum}. Increased dimensionality enables more compact circuit designs~\cite{ralph2007efficient,li2013geometry,Lanyon2008,liu2020optimal}, improves the efficiency of fault-tolerant quantum computation~\cite{bocharov2017factoring,babazadeh2017high,Muralidharan2017,campbell2014enhanced,howard2017application,campbell2012magic,menicucci2006universal,gu2009quantum,Amin2012}, enhances communication capacity~\cite{bechmann2000quantum,cortese2004holevo,dixon2012quantum,hu2018beating,Wang2015,hao2021entanglement,Ding2016,parigi2015storage,dkabrowski2018certification}, and increases robustness against noise and eavesdropping~\cite{cerf2002security,bouchard2017high,islam2017provably}. High-dimensional systems also constitute richer resources for quantum simulation~\cite{aaronson2011computational,hamilton2017gaussian,neeley2009emulation} and offer advantages in quantum metrology~\cite{lloyd2008enhanced,Chen:25,PhysRevApplied.21.034004}. Moreover, high-dimensional states exhibit stronger violations of Bell-type inequalities~\cite{collins2002bell,vertesi2010closing,dada2011experimental,collins2002bell}.

At the core of high-dimensional quantum technologies lie deterministic,
arbitrary multidimensional unitary gates capable of operating with high
fidelity. Realizing such gates, however, still requires substantial
effort to achieve both scalability and precision. Photons offer a particularly
versatile platform: their multiple accessible degrees of freedom—spatial,
temporal, and frequency modes—naturally support high-dimensional encodings.
Recent demonstrations of high-dimensional quantum gates in both
spatial~\cite{zhong2020quantum,carolan2015universal} and
temporal~\cite{madsen2022quantum} domains highlight this potential. Nevertheless, these approaches typically require $\sim m^{2}/2$ two-dimensional gate primitives to synthesize an $m$-dimensional unitary, demanding many optical components together with stringent phase stability and synchronization across large experimental setups. This overhead introduces significant loss and limits achievable fidelity, posing major challenges for further scaling and on-chip integration.

Spectral encoding offers an alternative route to implementing high-dimensional quantum gates. Electro-optic modulator based schemes, combined with pulse shapers, provide reconfigurability and flexibility but rely on active spectral modulation of the quantum field~\cite{Lukens:17,lu2019controlled,PhysRevLett.120.030502,Lu:18,PhysRevLett.125.120503,Lu:23,lu2022bayesian,PhysRevLett.129.230505,Lingaraju:22}. This introduces optical losses and constrains the achievable dimensionality due to the complex waveforms required for radio-frequency driving signals. Another promising approach employs a multi-output quantum pulse gate (QPG) based on dispersion-engineered sum-frequency generation (SFG) processes, enabling programmable frequency-bin interferometers~\cite{PRXQuantum.4.020306,PRXQuantum.5.040329,PhysRevResearch.6.L022040}. However, these systems are fundamentally limited to a maximal conversion efficiency (CE) of approximately 0.8 due to time-ordering effects~\cite{PhysRevA.90.030302,christ2013theory,Reddy:14,PRXQuantum.4.020306}, making them intrinsically non-deterministic. The optical losses reduce or even eliminate quantum advantage in practical applications~\cite{qi2020regimes,Chen:25}.

In this work, we propose a deterministic, universal, and fully programmable optical quantum gate for
high-dimensional systems based on a cavity-assisted SFG (CSFG)
process, achieving near-unity fidelity. The device implements an $M\times N$
truncated unitary transformation ($1\leq M< N$), or a full unitary when
$M=N$, on the input frequency-bin modes. With current state-of-the-art
technology, the attainable dimensionality can reach $M\times N\sim10^{4}$, and
$N$ may extend to $10^{3}$; even higher dimensions are feasible by employing
multiple pulse shapers. When combined with quantum light sources (both existing
and those introduced here), high-efficiency detection, and fast classical
feed-forward, this approach provides an efficient and scalable platform for
high-dimensional frequency-bin quantum processing, realized within a single
fiber-optic spatial mode that offers intrinsic phase stability and compatibility
with existing fiber-network infrastructures.

\begin{figure}[t]
    \centering
    \includegraphics[width=1\linewidth]{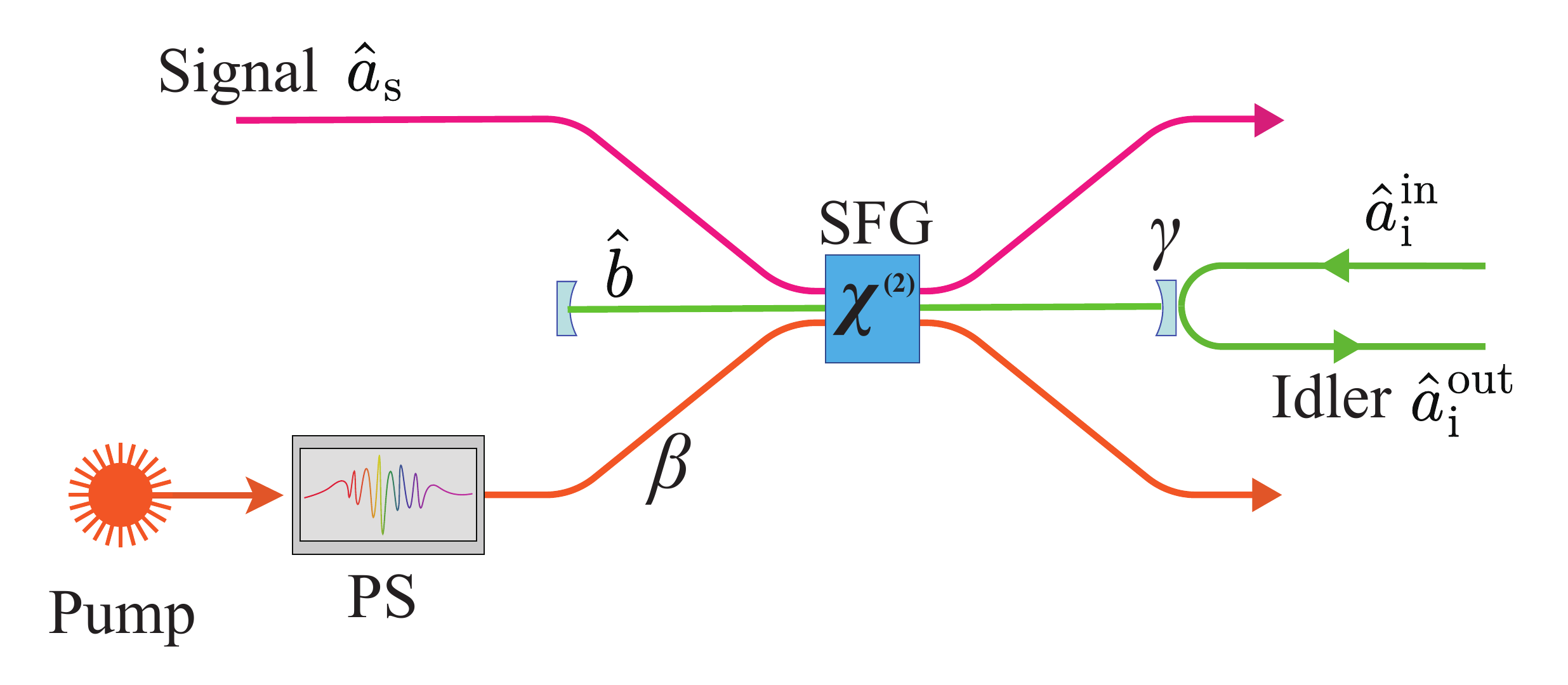}
    \caption{Schematic illustration of the $1\times N$ gate. PS: pulse shaper.}
    \label{fig:1}
\end{figure}
\section{Theoretical model}
\subsection{The analysis of the $1\times N$ gate}
We first implement the \(1\times N\) gate—also known as the QPG~\cite{Eckstein:11,PhysRevA.90.030302,Reddy:13,Reddy:14,Reddy:18}—and then naturally extend it to the general \(M\times N\) case. The \(1\times N\) gate consists of a strong pump pulse, spectrally shaped (via a Fourier-transform pulse shaper) to define the selected gate mode, together with an input signal field. When the signal’s temporal mode (TM) \cite{Raymer_2020,PhysRevX.5.041017} matches that of the pump, the two interact and up-convert the signal into a new idler frequency, whereas orthogonal TMs pass through essentially unchanged. A preliminary $1\times N$ gate was previously demonstrated in a task-specific
protocol~\cite{Chen:25}, whose validation relied on a particular TM structure.
Here, we develop a general, TM-independent theory and show that the same
mechanism enables universal and fully programmable gating. The overall concept is illustrated in Fig.~\ref{fig:1}.

All fields are assumed to occupy a single spatial mode (which can be realized using a waveguide) observed within a finite time window $T$. We take the fields to be one-dimensional and quasi-monochromatic \cite{ou2017quantum}. The strong, undepleted pump pulse has a normalized temporal profile $\beta(t)$, defined such that $\int_{-T/2}^{T/2} |\beta(t)|^{2}\, dt = 1$, and drives a $\chi^{(2)}$ nonlinear medium. In the SFG process, the pump field and the signal field $\hat{a}_{\mathrm{s}}$ interact to generate an idler field at their sum frequency. The idler field is resonantly enhanced by a cavity at frequency $\omega_{\mathrm{i},\mathrm{c}}$, with intracavity mode operator $\hat{b}$ coupled to the external field $\hat{a}_{\mathrm{i}}$. Assuming the pump and signal bandwidths are narrower than the cavity free spectral range (FSR) $\Omega_{\mathrm{FSR}}$, only a single resonance contributes. By appropriately choosing the carrier frequencies and polarizations, both energy-conservation and phase-matching conditions are satisfied.

In a rotating frame defined by the transformations
$\hat{a}_{j}(\omega_{n}) \rightarrow
\hat{a}_{j}(\omega_{n}) e^{-i\omega_{j,\mathrm{c}} t}$,
$\hat{b} \rightarrow \hat{b} e^{-i\omega_{\mathrm{i},\mathrm{c}} t}$, and
$\beta(t) \rightarrow \beta(t) e^{-i\omega_{\mathrm{p},\mathrm{c}} t}$,
where $j\in\{\mathrm{s},\mathrm{i}\}$ labels the signal ($\mathrm{s}$) and
idler ($\mathrm{i}$) fields, and the carrier frequencies satisfy the
energy-matching condition
 $\omega_{\mathrm{s},\mathrm{c}}+\omega_{\mathrm{p},\mathrm{c}}
=\omega_{\mathrm{i},\mathrm{c}}$, the system Hamiltonian can be written as (see Appendix \ref{1nda})
\bal
\hat{H}=\hat{H}_0+\hat{H}_1 ,
\label{Hamitonianallro2}
\eal
with
\bal
\hat{H}_0/\hbar &=
\sum_{j\in\{\mathrm{s},\mathrm{i}\}}\sum_{n}
\omega_{n}\,
\hat{a}^{\dagger}_{j}(\omega_{n})
\hat{a}_{j}(\omega_{n})
+ i\sqrt{\gamma}\!\left(
\hat{a}^{\dagger}_{\mathrm{i}}\hat{b}
-\hat{b}^{\dagger}\hat{a}_{\mathrm{i}}
\right), \\
\hat{H}_1/\hbar &=
-i\eta\!\left[
\hat{a}_{\mathrm{s}}\hat{b}^{\dagger}\beta(t)
-\hat{a}_{\mathrm{s}}^{\dagger}\hat{b}\beta^{*}(t)
\right].
\nonumber
\eal
Here
\(
\hat{a}_{j}
=
\sqrt{1/T}\sum_{n}\hat{a}_{j}(\omega_{n}),
\)
where $\omega_{n}=n\Delta\omega$ $(\Delta\omega=2\pi/T)$ label discrete
frequency bins.
The spatial integrations in the Hamiltonian have already been carried out. In $\hat{H}_{1}$, the resulting phase-matching factor
\(
C = \int_{0}^{L} e^{i\Delta k z}\, dz
\)
is treated as a constant. This approximation is justified by choosing the cavity linewidth and pump
bandwidth sufficiently small so that only frequency components with
$\Delta k L \ll 1$ acquire appreciable gain, whereas off--phase-matched
components are strongly suppressed.
The phase mismatch is
$\Delta k = k_{\mathrm{i}} - k_{\mathrm{s}} - k_{\mathrm{p}}$,
with $k_{\mathrm{s}}, k_{\mathrm{p}}, k_{\mathrm{i}}$ the signal, pump, and
idler wavevectors and $L$ the nonlinear-interaction length~\cite{Chen:25}. Finally, $\eta>0$ is the effective nonlinear coupling strength and $\gamma$ is the cavity external coupling rate. From Eq.~(\ref{Hamitonianallro2}), the Heisenberg–Langevin equation for the cavity mode $\hat{b}$ is
\be
\dot{\hat{b}}(t)=-\frac{\gamma}{2}\hat{b}(t)-\eta\hat{a}^{\mathrm{in}}_{\mathrm{s}}(t)\beta(t)-\frac{\eta^2}{2}\hat{b}(t)|\beta(t)|^2-\sqrt{\gamma}\hat{a}^{\mathrm{in}}_{\mathrm{i}}(t),
\label{ceq3}
\ee
with periodic boundary condition $\hat{b}(T/2)=\hat{b}(-T/2)$. Input/output operators are defined in the time domain as 
\(
\hat{a}^{\mathrm{in(out)}}_{j}(t)
=
(1/\sqrt{T})\sum_{n}
\hat{a}^{\mathrm{in(out)}}_{j}(\omega_{n})
\,\mathrm{e}^{-i\omega_{n} t},
\)
with
\(
\hat{a}^{\mathrm{in(out)}}_{j}(\omega_{n})
=
\hat{a}_{j}(\omega_{n},\mp T/2)\,
\mathrm{e}^{\mp i\omega_{n} T/2}.
\)

The solution of Eq. (\ref{ceq3}) 
reads
\begin{align}
\hat{b}(t)
=\sum_{j}\int_{-T/2}^{T/2} g'_j(t,t')\,
\hat{a}_{j}^{\text{in}}(t')\,dt',
\label{boutgc}
\end{align}
where the kernels are
\bal
g'_j(t,t')
&=-\left[
\frac{e^{-\gamma T/2-\eta^2/2}}
{1-e^{-\gamma T/2-\eta^2/2}}
+\Theta(t-t')
\right]\\&\times
h_j(t')
\exp\!\left[-\int_{t'}^{t}
\left(\frac{\gamma}{2}+\frac{\eta^{2}}{2}|\beta(t'')|^{2}\right)
dt''\right],
\label{g1a}
\eal
with
$h_{\text{s}}(t)=\eta\beta(t)$, $h_{\text{i}}(t)=\sqrt{\gamma}$
and \(\Theta\) the Heaviside step function. 
Using the cavity input–output relation
\be
\hat{a}_{\rm i}^{\rm out}(t)-\hat{a}_{\rm i}^{\rm in}(t)=\sqrt{\gamma}\hat{b}(t),
\label{inoutcaa}
\ee
together with Eq. (\ref{boutgc}), the idler output operator can be written directly as
\be
\hat{a}_{\rm i}^{\rm out}(t)=\sum_{j}\int_{-T/2}^{T/2} g_j(t,t')\,
\hat{a}_{j}^{\text{in}}(t')\,dt',
\label{inoutg}
\ee
where $g_{\rm s}(t,t')=\sqrt{\gamma} g'_{\rm s}(t,t')$ and $g_{\rm i}(t,t')=\sqrt{\gamma}g'_{\rm i}(t,t')+\delta(t,t')$.

To access the operating regime of interest, we consider the limit $\eta\sim\sqrt{\gamma T}\rightarrow 0$  ($\gamma/\Delta\omega\rightarrow 0$). In this regime, Eq. (\ref{g1a}) reduces to
\be
g'_j(t,t')=-2 h_j(t')/(\gamma T+\eta^2),
\ee
The idler output field in the frequency domain then follows as
\begin{align}
\hat{a}_{\text{i}}^{\text{out}}(\omega_{n})
=
\begin{cases}
\mu'_0\hat{A}_{\text{s}}^{\text{in}}(0)
+\nu'_0\hat{a}_{\text{i}}^{\text{in}}(\omega_0),
& n=0,\\[6pt]
\hat{a}_{\text{i}}^{\text{in}}(\omega_{n}),
& n\neq 0.
\end{cases}
\label{solfm}
\end{align} 
where \[\mu'_0=-\frac{2\eta\sqrt{\gamma T}}{\gamma T+\eta^2},\quad\nu'_0=\frac{\eta^2-\gamma T}{\gamma T+\eta^2},\]
and the TM $\hat{A}^{\mathrm{in}}_{\mathrm{s}}(0)=\sum_m\beta(\omega_{-m})\hat{a}^{\mathrm{in}}_{\mathrm{s}}(\omega_{m})$. In particular, when $\eta=\sqrt{\gamma T}$, the output simplifies to
\begin{align}
	\hat{a}_{\text{i}}^{\text{out}}(\omega_{n})
	=
	\begin{cases}
		-\hat{A}_{\text{s}}^{\text{in}}(0), & n=0,\\[3pt]
		\hat{a}_{\text{i}}^{\text{in}}(\omega_{n}), & n\neq 0,
	\end{cases}
	\label{fullca}
\end{align}
and in this regime both the fidelity and CE approach unity \cite{lu2019controlled,Reddy:13}. 
The fidelity quantifies how closely the implemented transformation matches 
the ideal one, while the CE gives the probability that the target mode 
$\hat{A}^{\mathrm{in}}_{\mathrm{s}}(0)$ is successfully converted.
\begin{figure}[t]
	\centering
	\includegraphics[width=\linewidth]{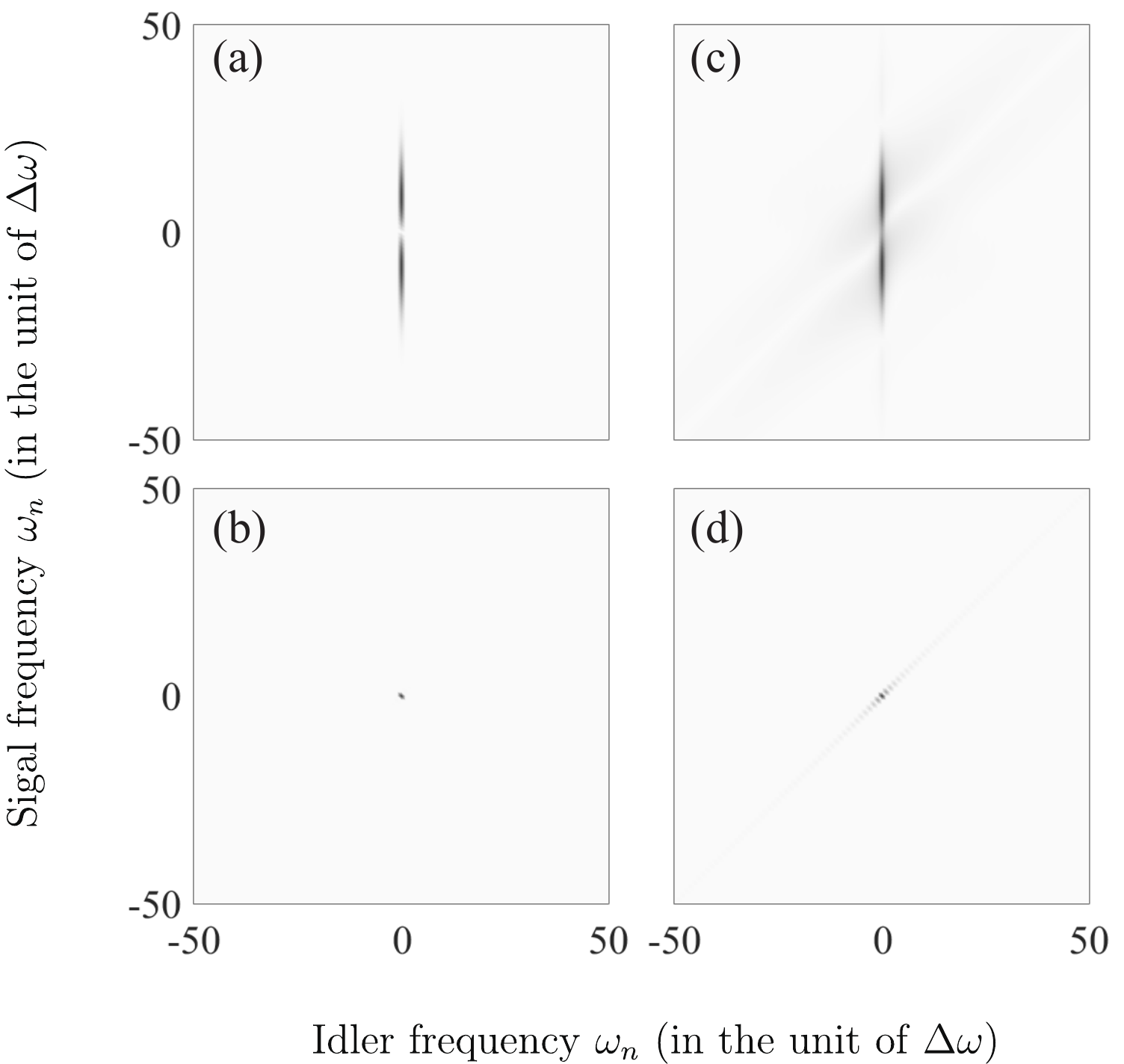}
	\caption{Amplitude of the transfer function $\tilde{g}_{\mathrm{s}}(\omega_{n},\omega_{m})$ for the $1\times N$ gate. For numerical tractability, the calculation is performed using 101 discrete frequency-bin modes. Panels (a) and (b) correspond to $\gamma/\Delta\omega=\eta^{2}/2\pi=0.01$, while panels (c) and (d) correspond to $\gamma/\Delta\omega=\eta^{2}/2\pi=0.5$. In panels (a) and (c), the pump field is encoded in a second-order Hermite–Gaussian mode, whereas in panels (b) and (d) it is encoded in the single-frequency-bin mode [$\beta(t)=1/\sqrt{T}$].}
	\label{fig:tr}
\end{figure}
\subsubsection{Fidelity and conversion efficiency}
To assess the gate’s performance away from the asymptotic limit, we evaluate its 
fidelity and CE in the regime of small but nonzero $\gamma/\Delta\omega$.
The idler output in Eq.~(\ref{inoutg}) can be expressed in the frequency basis as
\be
\hat{a}_{\rm i}^{\rm out}(\omega_{n})=\sum_{j,m} \tilde{g}_{j}(\omega_{n},\omega_{m})\,
\hat{a}_{j}^{\text{in}}(\omega_m),
\label{inoutgg}
\ee
where
\be
	\tilde{g}_{j}(\omega_n,\omega_m)
	=\frac{1}{T}
	\iint_{-T/2}^{T/2} dt
	dt'\,
	e^{i\omega_n t}\,
	g_j(t,t')\,
	e^{-i\omega_{m} t'}.
\label{kernelr}
\ee
Although this expression formally defines an $N\times N$ linear transformation $\tilde{g}_{\rm s}$ acting on the signal’s frequency-bin modes, the resulting matrix is highly sparse: only the resonant idler output mode acquires an appreciable amplitude, as shown in Fig. \ref{fig:tr}. Consequently, the device operates effectively as a $1\times N$ gate in this regime. 
We consider the full $N\times N$ matrix fidelity, taking the ideal map to be
$\tilde{g}_{\rm s}^{\rm ideal}$, defined by
$\tilde{g}_{\rm s}^{\rm ideal}(0,m)=\beta(m)$, with all other entries set to
zero.
Throughout this paper, we write $f(m)$ and $f(n,m)$ in place of
$f(\omega_m)$ and $f(\omega_n,\omega_m)$ when referring to vector or matrix
elements. The full-matrix (FM) fidelity is defined as the Hilbert--Schmidt overlap
of the two map matrices,
\bal
\mathcal{F}_{i}^{\rm FM}&=\frac{|{\rm Tr}(\tilde{g}_{\rm s}^{\dagger}\tilde{g}_{\rm s}^{\rm ideal})|^2}{{\rm Tr}(g_{\rm s}^{\dagger}\tilde{g}_{\rm s}){\rm Tr}(\tilde{g}_{\rm s}^{{\rm ideal}\dagger}\tilde{g}_{\rm s}^{\rm ideal})}
\\&=
\frac{
	\displaystyle 
	\frac{1}{T}\,
	\bigg|
	\int_{-T/2}^{T/2}\! dt
	\int_{-T/2}^{T/2}\! du\;
	g_{\rm s}(t,u)\,\beta^{*}(u)
	\bigg|^{2}
}{
	\displaystyle 
		\iint_{-T/2}^{T/2}
	dt\,du\;
	|g_{\rm s}(t,u)|^{2}
}.
\label{fidem1}
\eal
where we have used $ \int_{-T/2}^{T/2} du \; |\beta(u)|^{2}=1$. Correspondingly, the CE is
\bal
\mathcal{C}_{e}^{\rm FM}&=\frac{|{\rm Tr}(\tilde{g}_{\rm s}^{\dagger}\tilde{g}_{\rm s}^{\rm ideal})|^2}{\big[{\rm Tr}(\tilde{g}_{\rm s}^{{\rm ideal}\dagger}\tilde{g}_{\rm s}^{\rm ideal})\big]^2}
\\&=
\frac{1}{T}\,
\bigg|
\int_{-T/2}^{T/2}\! dt
\int_{-T/2}^{T/2}\! du\;
g_{\rm s}(t,u)\,\beta^{*}(u)
\bigg|^{2}.
\label{cem1}
\eal
However, when the device is regarded as a $1\times N$ gate, it is typically followed by a single detector, as discussed in a later section. In this configuration, the effective transformation seen by a single detector becomes measurement-dependent; for instance, photon counting (PC) with limited
spectral resolution is sensitive only to the total photon number
across all frequency-bin modes and is insensitive to their relative phases.
As a result, the corresponding measurement-dependent fidelity and CE are always greater than or equal to their FM counterparts (see Appendix~\ref{1nfice}). Consequently, the FM fidelity and CE remain appropriate figures of merit: they characterize the gate’s performance when coherently cascaded with subsequent operations (e.g., another logic gate) and simultaneously provide a conservative lower bound for scenarios in which the outputs are measured directly by a single detector.

\begin{figure}[t]
	\centering
	\includegraphics[width=\linewidth]{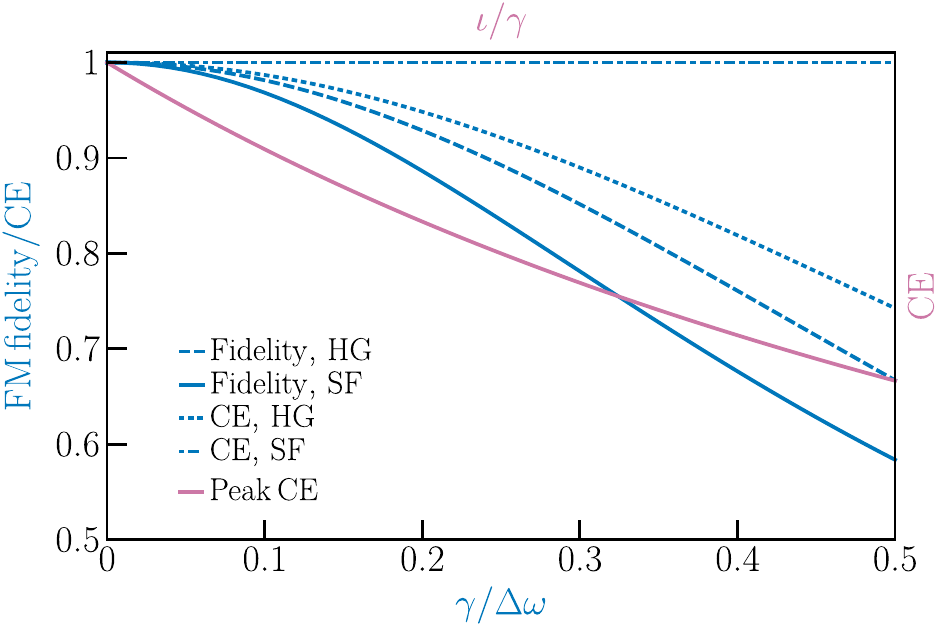}
	\caption{
		FM fidelity and CE of the effective $1\times N$ gate versus 
		$\gamma/\Delta\omega$, evaluated under the condition $\eta=\sqrt{\gamma T}$, 
		for a second-order HG pump and an SF pump. 
		The pink curve shows the peak achievable CE as a function of $\iota/\gamma$ 
		in the limit $\eta=\sqrt{(\gamma+\iota)T}\to 0$.  
		}
	\label{fig:2}
\end{figure}

Fig.~\ref{fig:2} shows the FM fidelity and CE of the $1\times N$ gate for pump spectra given by a second-order Hermite–Gaussian (HG) mode and a single-frequency-bin (SF) mode [$\beta(t)=1/\sqrt{T}$], plotted as functions of $\gamma/\Delta\omega$ under the constraint $\eta=\sqrt{\gamma T}$. The degradation of gate performance with increasing $\gamma/\Delta\omega$ is consistent with the behavior observed in Fig.~\ref{fig:tr}, where signal–idler frequency correlations become stronger as $\gamma/\Delta\omega$ increases. The CE for the SF mode remains unity, as follows from the solution of Eq.~(\ref{ceq3}) for a flat pump intensity $|\beta(t)|^{2}=1/T$ (see Appendix~\ref{cmss}).

\subsubsection{Effect of internal loss}
To incorporate internal losses, we model coupling to an ancillary bath mode $\hat{d}^{\mathrm{in}}$ with loss rate $\iota$. In this case, Eq.~(\ref{ceq3}) becomes
\bal
\dot{\hat{b}}(t)
&=
-\frac{\gamma}{2}\hat{b}(t)
-\frac{\iota}{2}\hat{b}(t)
-\eta\,\hat{a}_{\text{s}}^{\text{in}}(t)\beta(t)
-\frac{\eta^{2}}{2}\hat{b}(t)|\beta(t)|^{2}\\&
-\sqrt{\gamma}\,\hat{a}_{\text{i}}^{\text{in}}(t)
-\sqrt{\iota}\,\hat{d}^{\text{in}}(t).
\label{ceqloss}
\eal
Applying the same method used for solving Eq.~(\ref{ceq3}) in the preceding section, 
and taking the limit $\eta \sim \sqrt{(\gamma+\iota)T}\rightarrow 0$, 
the idler output in the frequency domain is found to be
\begin{align}
	\hat{a}_{\mathrm{i}}^{\mathrm{out}}(\omega_{n})
	=
	\begin{cases}
		\mu''_{0}\,\hat{A}_{\mathrm{s}}^{\mathrm{in}}(0)
		+\nu''_{0}\,\hat{a}_{\mathrm{i}}^{\mathrm{in}}(\omega_{0})
		+\upsilon''_{0}\,\hat{d}^{\mathrm{in}}(\omega_{0}),
		& n=0,\\[4pt]
		\hat{a}_{\mathrm{i}}^{\mathrm{in}}(\omega_{n}),
		& n\neq 0,
	\end{cases}
\end{align}
with coefficients
\begin{align}
\mu''_{0} &= -\,\frac{2\eta\sqrt{\gamma T}}{\gamma T + \iota T + \eta^{2}}, \nonumber\\[3pt]
\nu''_{0} &= \frac{\eta^{2} - \gamma T + \iota T}{\gamma T + \iota T + \eta^{2}}, \nonumber\\[3pt]
\upsilon''_{0} &= -\,\frac{2\sqrt{\gamma\iota}}{\gamma T + \iota T + \eta^{2}}. 
\nonumber
\end{align}
 The CE, given by $|\mu''_{0}|^{2}$, reaches its maximum value $1/(1+\iota/\gamma)$ at $\eta = \sqrt{(\gamma+\iota)T}$, as shown in Fig.~\ref{fig:2}, indicating that internal losses reduce the achievable peak CE. An ideal deterministic gate requires $\iota = 0$, whereas in practice a near-deterministic gate can be realized when the relative internal-loss ratio $\iota/\gamma$ is sufficiently small.

\subsection{The $M\times N$ gate}
The $1\times N$ quantum gate based on the CSFG naturally extends to an $M\times N$ gate. Let the cavity resonances be $\omega_{\mathrm{i},\mathrm{c},m}=\omega_{\mathrm{i},\mathrm{c}}+m\Omega_{\mathrm{FSR}}$, and denote the corresponding external idler modes by
\(
\hat{a}_{\mathrm{i},m}
=
\sqrt{1/T}\sum_{n}\hat{a}_{\mathrm{i},m}(\omega_{n}),
\)
each with an identical bandwidth smaller than $\Omega_{\mathrm{FSR}}$ and coupled to intracavity mode $\hat{b}_m$. The signal field $\hat{a}_{\mathrm{s}}$ is assumed to have the same bandwidth. The pump field is taken as $\beta(t)=\sum_{m}\beta_{m}(t)\,{\rm e}^{-i\omega_{\mathrm{p},\mathrm{c},m} t}$, where each tone $\beta_{m}(t)\,{\rm e}^{-i\omega_{\mathrm{p},\mathrm{c},m} t}$ has a bandwidth matched to the signal and idler fields and is centered at  $\omega_{\mathrm{p},\mathrm{c},m}=\omega_{\mathrm{p},\mathrm{c}}+m\Omega_{\mathrm{FSR}}$. The pump envelopes $\{\beta_m(t)\}$ are mutually orthogonal.

In the rotating frame defined by
$\hat{a}_{\mathrm{s}}(\omega_{n}) \rightarrow
\hat{a}_{\mathrm{s}}(\omega_{n}) e^{-i\omega_{\mathrm{s},\mathrm{c}} t}$,
$\hat{a}_{\mathrm{i},m}(\omega_{n}) \rightarrow
\hat{a}_{\mathrm{i},m}(\omega_{n}) e^{-i\omega_{\mathrm{i},\mathrm{c},m} t}$,
and $\hat{b}_m \rightarrow \hat{b}_m e^{-i\omega_{\mathrm{i},\mathrm{c},m} t}$,
the Langevin equations governing the cavity modes take the form
\cite{PhysRevA.43.543}
\bal
\dot{\hat{b}}_k(t)
&=-\frac{\gamma}{2}\hat{b}_k(t)
-\eta\,\hat{a}_{\mathrm{s}}^{\mathrm{in}}(t)\beta_k(t)
\\
&\quad
-\frac{\eta^2}{2}\sum_m \hat{b}_m(t)\beta_m^{*}(t)\beta_k(t)
-\sqrt{\gamma}\hat{a}_{\mathrm{i},k}^{\mathrm{in}}(t),
\label{ceqmul}
\eal
where rapidly rotating terms proportional to
$e^{-i m \Omega_{\mathrm{FSR}} t}$ ($m\neq 0$) have been neglected.
Together with the cavity input--output relations
\be
\hat{a}_{\mathrm{i},k}^{\mathrm{out}}(t)
-\hat{a}_{\mathrm{i},k}^{\mathrm{in}}(t)
=\sqrt{\gamma}\hat{b}_k(t),
\label{inoutcaamch}
\ee
Eq.~(\ref{ceqmul}) admits an analytic solution in the limit
$\eta\sim\sqrt{\gamma T}\to 0$ ( $\gamma/\Delta\omega\to 0$),
under which the dynamics decomposes into independent resonance channels (see Appendix~\ref{mnda}).
In this regime, the idler output at each resonance is given by
\begin{align}
	\hat{a}_{\mathrm{i},k}^{\mathrm{out}}(\omega_{n})
	=
	\begin{cases}
		\mu'_0\,\hat{A}_{\mathrm{s},k}^{\mathrm{in}}(0)
		+\nu'_0\,\hat{a}_{\mathrm{i},k}^{\mathrm{in}}(\omega_0),
		& n=0,\\[6pt]
		\hat{a}_{\mathrm{i},k}^{\mathrm{in}}(\omega_{n}),
		& n\neq 0,
	\end{cases}
	\label{solfmmn}
\end{align}
where
\bal
\hat{A}_{\mathrm{s},k}^{\mathrm{in}}(0)
&=\int_{-T/2}^{T/2}\! dt'\,
\beta_k(t')\,\hat{a}_{\mathrm{s}}^{\mathrm{in}}(t')
\nonumber\\
&=\sum_{n}
\beta_k(\omega_{-n})\,
\hat{a}_{\mathrm{s}}^{\mathrm{in}}(\omega_{n}),
\eal
which is the direct multi-channel generalization of Eq.~(\ref{solfm}). 
In the special case $\eta=\sqrt{\gamma T}$, the outputs simplify to
\be
\hat{a}_{\mathrm{i},k}^{\mathrm{out}}(\omega_{n})
=
\begin{cases}
	-\hat{A}_{\mathrm{s},k}^{\mathrm{in}}(0), & n=0,\\[3pt]
	\hat{a}_{\mathrm{i},k}^{\mathrm{in}}(\omega_{n}), & n\neq 0.
\end{cases}
\ee
This implements an $M\times N$ truncated-unitary transformation (unitary when $M=N$) on the frequency-bin modes \cite{PRXQuantum.5.040329}, yielding the idler output \be
\hat{a}_{\mathrm{i},m}^{\mathrm{out}}(\omega_{0})=\sum_{l} U_{ml}\hat{a}^{\mathrm{in}}_{\mathrm{s}}(\omega_{l}),
\ee
where $U_{ml}=-\beta_{m}(\omega_{-l})$. In this asymptotic limit, both the fidelity and the CE approach unity.
\begin{figure}[t]
	\centering
	\includegraphics[width=\linewidth]{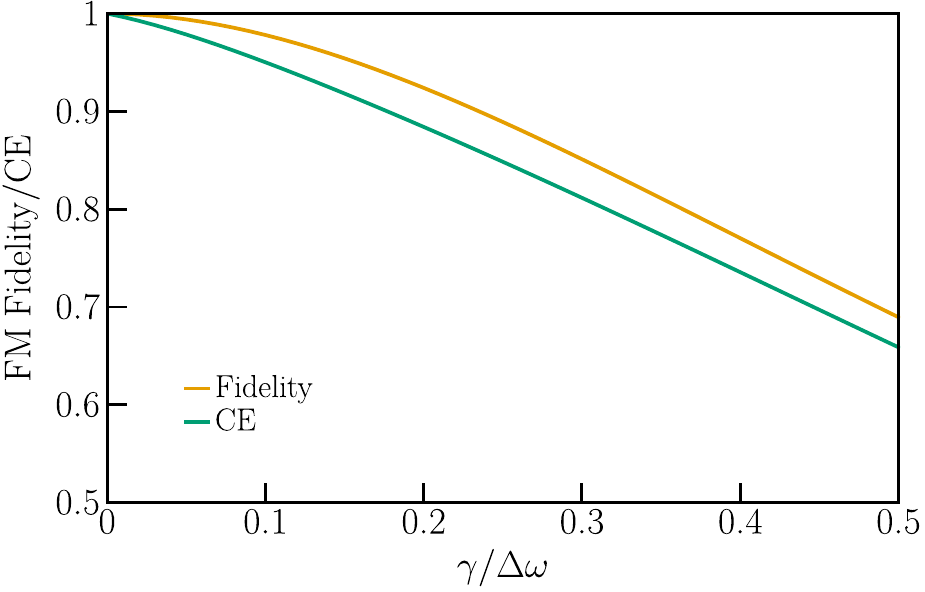}
	\caption{
		FM fidelity and CE, for a $101\times101$ gate versus $\gamma/\Delta\omega$, evaluated with respect to the ideal identity transformation under $\eta=\sqrt{\gamma T}$, using the pump 
		$\beta(t)=-(1/\sqrt{T})\sum_{m} e^{-i(\omega_{\mathrm{p},\mathrm{c},m}-m\Delta\omega)t}$. 
	}
	\label{fig:22}
\end{figure}
\subsubsection{Fidelity and conversion efficiency}
We evaluate the fidelity and CE of the $M\times N$ gate in the regime of small but nonzero $\gamma/\Delta\omega$. In this regime, the solutions of Eqs.~(\ref{ceqmul}) can be written in the compact form (see Appendix~\ref{mnficesec})
\bal
\vec a_{\rm i}^{\,\rm out}(t)
&=
\int_{-T/2}^{T/2}du\;\vec G_{\rm s}(t,u)\,\hat a_{\rm s}^{\,\rm in}(u)
\\&+
\int_{-T/2}^{T/2}du\;\mathbb{G}_{\rm i}(t,u)\,\vec a_{\rm i}^{\,\rm in}(u),
\eal
where $\vec a_{\rm i}^{\,\rm in(out)}(t)
=
\{\hat a_{{\rm i},k}^{\,\rm in(out)}(t)\}^T$ and $\vec G_{\rm s}(t,u)=\{g_{{\rm s},k}(t,u)\}^T$ are $M$-component vectors and $\mathbb{G}_{\rm i}$ is an $M\times M$ matrix. The corresponding frequency-domain expression is
\bal
\vec a_{\rm i}^{\,\rm out}(\omega_n)
&=
\sum_m \;\vec{\tilde{G}}_{\rm s}(\omega_n,\omega_m)\,\hat a_{\rm s}^{\,\rm in}(\omega_m)
\\&+
\sum_m \;\tilde{\mathbb{G}}_{\rm i}(\omega_n,\omega_m)\,\vec a_{\rm i}^{\,\rm in}(\omega_m),
\label{mnf}
\eal
where $\vec{\tilde{G}}_{\rm s}(\omega_n,\omega_m)=\{\tilde{g}_{{\rm s},k}(\omega_n,\omega_m)\}^T$. For each channel $k$, the functions $\tilde{g}_{{\rm s},k}(\omega_n,\omega_m)$ and $g_{{\rm s},k}(t,u)$ satisfy the same single-channel relation given in Eq.~(\ref{kernelr}).

Similar to the $1\times N$ gate case, this expression formally defines an $MN\times N$ linear transformation $\vec{\tilde{G}}_{\rm s}$ acting on the signal’s frequency-bin modes. Owing to the resonant structure of the interaction, this transformation is highly sparse. We take the ideal transformation to be $\vec{\tilde{G}}_{\rm s}^{\rm ideal}(0,m)=\vec{\beta}(m)$, where $\vec{\beta}=\{\beta_k\}^{T}$, with all other matrix elements set to zero. The FM fidelity of the $M\times N$ gate is then evaluated as the Hilbert–Schmidt overlap between the actual and ideal transformations,
\bal
\mathcal{F}_i^{\mathrm{FM}}
&=
\frac{\big|\mathrm{Tr}(\tilde{G}_{\rm s}^{\dagger}\tilde{G}_{\rm s}^{\rm ideal})\big|^{2}}
{\mathrm{Tr}(\tilde{G}_{\rm s}^{\dagger}\tilde{G}_{\rm s})\,
	\mathrm{Tr}({\tilde{G}_{\rm s}}^{{\rm ideal}\dagger}\tilde{G}_{\rm s}^{\rm ideal}) }
\\&=
\frac{\displaystyle 
	\left|
	\iint_{-T/2}^{T/2} dt\;du\;
	\vec{\beta}^{\,\dagger}(u)\,G_{\rm s}(t,u)
	\right|^{2}}
{\displaystyle 
	MT
	\iint_{-T/2}^{T/2} dt\,du\;
	\big\|G_{\rm s}(t,u)\big\|_{2}^{2}
},
\label{mnfmfide}
\eal
where we have used $\int_{-T/2}^{T/2} du\;\|\vec\beta(u)\|_{2}^{2}=M$, and $\|\cdot\|_{2}$ denotes the Euclidean ($\ell^{2}$) norm, defined for any vector $x$ as $\|x\|_{2}=\sqrt{x^{\dagger}x}$. The corresponding FM CE is given by
\bal
\mathcal{C}_e^{\mathrm{FM}}
&=
\frac{\big|\mathrm{Tr}(\tilde{G}_{\rm s}^{\dagger}\tilde{G}_{\rm s}^{\rm ideal})\big|^{2}}
{\big[\mathrm{Tr}({\tilde{G}_{\rm s}}^{{\rm ideal}\dagger}\tilde{G}_{\rm s}^{\rm ideal})\big]^2}
\\&=
\frac{1}{M^2 T}
\left|
\iint_{-T/2}^{T/2} dt\;du\;
\vec\beta^{\,\dagger}(u)\,\vec G_{\rm s}(t,u)
\right|^{2}.
\label{fmcenn}
\eal

Fig.~\ref{fig:22} shows the FM fidelity and CE of a $101\times101$ gate as functions of $\gamma/\Delta\omega$, evaluated with respect to the ideal identity transformation. The results are obtained using the pump field $\beta(t)=-(1/\sqrt{T})\sum_{m}{\rm e}^{-i(\omega_{\mathrm{p},\mathrm{c},m}-m\Delta\omega)t}$ with the coupling constrained to $\eta=\sqrt{\gamma T}$. When placed before a wavelength-division demultiplexing (WDM) stage, this identity gate enhances frequency separation and enables demultiplexing at denser channel spacings. Measurement, such as PC or homodyne detection (HD), generically enhances both the fidelity and the CE (see Appendix~\ref{mnficesec}). Consequently, the FM fidelity and CE constitute a conservative lower bound for configurations in which the gate output is fully measured, as well as for hybrid scenarios where only a subset of outputs is measured and the remaining modes are coherently routed to subsequent logic gates. As shown in Fig.~\ref{fig:22}, the gate performance exhibits the same overall degradation with increasing $\gamma/\Delta\omega$ as in the $1\times N$ gate case.

\section{Schemes for high-dimensional quantum processing}
Combined with nonclassical sources, measurement, and fast feed-forward, the quantum gate provides an effective and scalable platform for high-dimensional frequency-bin quantum processing, supporting applications in quantum computation, simulation, communication, and sensing. We first describe compatible nonclassical sources and then outline several
representative schemes enabled by the gate.

A continuous-wave pump with linewidth $\ll \Delta\omega$ drives spontaneous parametric down-conversion (SPDC) within the observation window $T$, naturally generating nonclassical states in discrete frequency-bin modes that are directly compatible with the gate \cite{Chen:25}. A nondegenerate SPDC source yields $M$ entangled signal–idler mode pairs in two-mode squeezed-vacuum (TMSV) states with uniform squeezing across modes, while a degenerate source generates $M$ single-mode squeezed-vacuum (SMSV) states with identical squeezing. The nondegenerate source, operated in the weak-interaction regime, additionally
enables heralded single photons distributed over $M$ frequency bins.
An alternative source is SPDC in a nonlinear resonator, which produces analogous squeezed, entangled, or heralded single-photon states directly in frequency-comb-bin modes \cite{ikuta2019frequency,kues2019quantum,jaramillo2017persistent}. Note that while we focus on SPDC, the same formalism applies to stimulated parametric down-conversion, yielding displaced squeezed states; this is optional for the schemes considered below.

\begin{figure}[t]
    \centering
 \includegraphics[width=\linewidth]{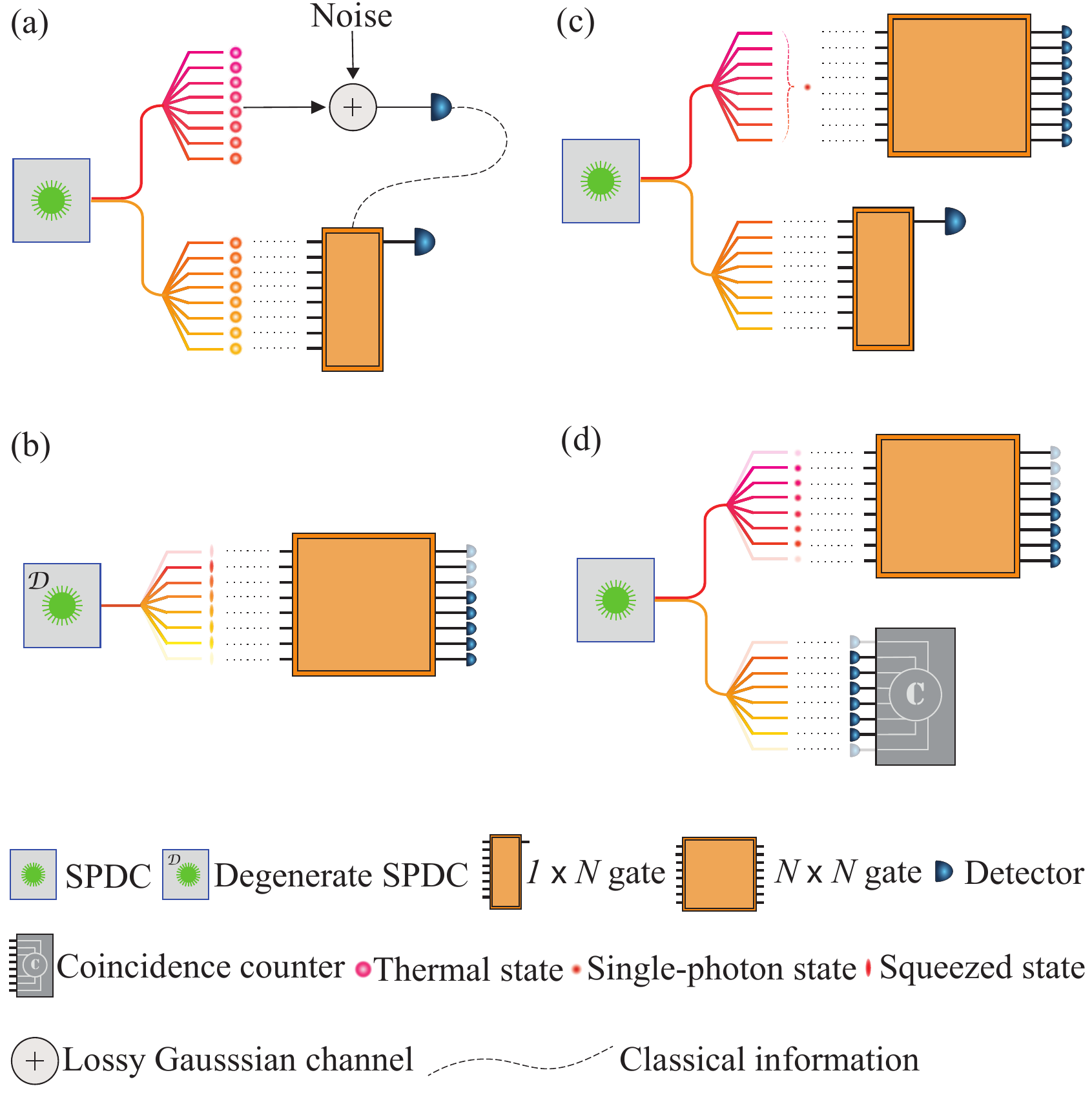}
    \caption{Representative schemes for high-dimensional quantum processing using the 
programmable frequency-bin gate.}
    \label{fig:3}
\end{figure}

Fig.~\ref{fig:3}(a) shows a scheme in which the signal modes of $N$ entangled pairs pass through a phase-shifted thermal-loss channel (e.g., target reflection or information encoding) and are then heterodyne-detected. The heterodyne outcomes are used to program a $1\times N$ gate acting on the idler modes, followed by PC. This realizes the correlation-to-displacement conversion protocol, enabling near-optimal performance in quantum illumination, phase sensing, and communication \cite{PhysRevA.107.062405,Chen:25,PhysRevApplied.20.014030,PhysRevApplied.21.034004,PhysRevApplied.20.024030}. Fig.~\ref{fig:3}(b) depicts a scheme in which $M$ ($M\leq N$) input SMSV states are processed by a fully programmable $N\times N$ gate, followed by WDM and PC or HD on a subset or all of the outputs. This architecture underlies applications such as Gaussian boson sampling \cite{hamilton2017gaussian} and measurement-based continuous-variable quantum computation (e.g., the generation of Gottesman–Kitaev–Preskill qubits) \cite{takase2023gottesman,tzitrin2020progress}.

The programmable $N\times N$ quantum gate also operates at the single-photon level.
Fig.~\ref{fig:3}(c) illustrates a programmable heralded multimode
single-photon (qudit) source prepared in the TM $\beta^{*}(t)$.
The state is obtained by heralding the signal output of a nondegenerate SPDC
source following spectral mode selection via a $1\times N$ gate pumped with
$\beta(t)$.
The resulting qudit state can be directly interfaced with an $N\times N$ gate for applications such as high-dimensional quantum key distribution~\cite{cerf2002security,khodadad2025frequency}. Fig.~\ref{fig:3}(d) shows the generation of $M$ ($M\leq N$) independent single-photon sources,
$\bigotimes_{n=-(M-1)/2}^{(M-1)/2}\ket{1}_n$,
each occupying a distinct frequency-bin mode, achieved by coincident heralding of the corresponding $M$ signal modes from a nondegenerate SPDC source. For WDM with dense channel spacing, the previously discussed $M\times M$ identity gate may be employed before the heralding detectors to enhance frequency separation. When combined with a programmable $N\times N$ gate, this configuration supports applications in boson sampling~\cite{carolan2015universal} and high-dimensional photonic quantum computation (e.g., the generation of high-dimensional GHZ states)~\cite{knill2001scheme,paesani2021scheme,carolan2015universal}.
In the single-photon regime, the programmable $N\times N$ gate can also operate as a high-dimensional Bell-state analyzer that enables entanglement swapping between remote matter qudits and boosts entanglement distribution rates by exploiting enlarged mode dimensionality \cite{sangouard2011quantum,chakraborty2025towards,sinclair2014spectral,ding2016high}.

\section{Scaling and limitations}
The dimensionality of the quantum gate is primarily limited by the nonlinear 
interaction bandwidth and the spectral resolution of the pump shaper.  
Although SFG can in principle support bandwidths of $1$--$10$~THz~\cite{ou2007multi}, 
faithful mode mapping requires the phase-matching response to remain flat 
across the pump bandwidth (so that the pump spectrum maps to the converted TM 
without distortion), reducing the usable signal bandwidth to 
$\Omega_{\rm s}\sim0.1$--$1$~THz (see Appendix \ref{1nda})~\cite{Chen:25}.  Taking $\Omega_{\rm s}=1$~THz and a commercial 
Fourier-transform pulse shaper with resolution $R\sim1$~GHz yields  
$N=\Omega_{\rm s}/R\sim10^3$ accessible frequency bins.  
For an $M\times N$ gate, the total dimensionality is further limited by the 
programmable bandwidth of the shaper, which currently supports on the order of 
$M\times N\sim 10^4$ modes. The dimensionality can be further extended by employing multiple pulse shapers. 
While cavity parameters (linewidth and FSR) affect the gate—high fidelity
and CE require $\gamma/\Delta\omega \lesssim 0.01$ and $\iota/\gamma \lesssim 0.01$ (Fig.~\ref{fig:2} and~\ref{fig:22}), and the
FSR bounds the usable signal bandwidth—current resonator technology
(sub-MHz linewidths and finesse up to $10^{6}$) readily accommodates the
dimensionalities considered here.

On the source side, an SPDC process pumped by a narrow-linewidth
continuous-wave laser serves as the time-reversed analog of CSFG, so its
phase-matching bandwidth naturally matches the mode-bandwidth requirement
of the $M\times N$ gate. Although the pump linewidth can, in principle,
influence fidelity and scalability of the system~\cite{PRXQuantum.5.040329}, modern
single-frequency lasers provide ultranarrow linewidths~\cite{bai2022comprehensive},
so the photon source does not impose a practical scalability constraint.

A practical constraint of the proposed architecture concerns the cascading of multiple CSFG-based gates. In particular, an $N\times N$ gate generally cannot be directly followed by another CSFG-based $M\times N$ gate, because the output frequency-bin spacing—set by the cavity resonances—is typically too large. Consequently, the gate is most naturally employed in configurations where either all output modes are directly measured or left in the vacuum state, or where only a subset of output modes is measured or left vacuum while a limited number of remaining modes are routed into subsequent low-dimensional frequency-bin or spatial-mode gates (e.g., for the generation of high-dimensional GHZ states~\cite{paesani2021scheme}).

\section{Conclusion}
We have proposed a deterministic, universal, and fully programmable
frequency-bin quantum gate based on CSFG. In the asymptotic regime
$\eta=\sqrt{\gamma T}\to 0$ ($\gamma/\Delta\omega\to 0$),
the device implements an $M\times N$ truncated-unitary transformation—
becoming fully unitary for $M=N$—with unity fidelity and conversion
efficiency. Away from this limit, we quantify the gate performance using
the FM fidelity and conversion efficiency as conservative benchmarks
for practical operating regimes, and show that the performance degrades
smoothly and predictably as $\gamma/\Delta\omega$ increases. We further
show that internal cavity loss reduces the achievable peak conversion
efficiency. Nevertheless, with state-of-the-art experimental parameters,
near-unit fidelity and near-deterministic operation are fully achievable
in practice.

Beyond the gate primitive itself, we have outlined representative
architectures that combine CSFG-based quantum gates with compatible
SPDC sources, high-efficiency detection, and fast feed-forward. These
architectures support programmable processing of multimode Gaussian
states for applications such as Gaussian boson sampling and
measurement-based continuous-variable quantum computation. They also
enable heralded preparation of multimode single-photon qudits and
multi-photon states, which can be deterministically routed
and transformed by an $N\times N$ programmable frequency-bin processor,
supporting high-dimensional quantum key distribution and photonic
quantum computation.

Finally, we discussed the principal scaling considerations and showed
that, with current technology, the attainable dimensionality of the
$M\times N$ quantum gate can reach $M\times N\sim10^{4}$, with $N$
extending up to $10^{3}$. Even higher dimensionalities are feasible by
employing multiple pulse shapers. Taken together, our results establish
CSFG as a scalable and low-loss approach that provides intrinsic phase
stability and fiber compatibility, offering a realistic platform for
high-dimensional frequency-bin quantum processing across quantum
computation, communication, sensing, and related photonic quantum
technologies.

\appendix
\section{Detailed analysis of the $1\times N$ gate}
\label{1nda}
In this section, we present a detailed analysis of the $1\times N$ gate based on the CSFG process. The total Hamiltonian reads
\bal
\hat{H}/\hbar&=\omega_{\rm i,c}\hat{b}^{\dagger}\hat{b}+\sum_{j,n}  [(\omega_{n}+\omega_{j,{\rm c}})\hat{a}^{\dagger}_j(\omega_{n})\hat{a}_j(\omega_{n})]\\&+i\sqrt{\gamma}\bigl(\hat{a}^{\dagger}_{\text{i}}\hat{b}-\hat{b}^{\dagger}\hat{a}_{\text{i}}\bigr)-i\eta\bigl[\hat{a}_{\text{s}}\hat{b}^{\dagger}\beta(t)-\hat{a}_{\text{s}}^{\dagger}\hat{b}\beta^{*}(t)\bigr].
\label{Hamitonianall}
\eal
Transforming to the rotating frame $\hat{a}_{j}(\omega_{n})\rightarrow\hat{a}_{j}(\omega_{n}){\rm e}^{-i\omega_{j,{\rm c}} t}$, $\hat{b}\rightarrow\hat{b}{\rm e}^{-i\omega_{\rm i,c} t}$, $\beta(t)\rightarrow \beta(t)e^{-i\omega_{\text{p,c}} t}$ with $\omega_{\text{s,c}}+\omega_{\text{p,c}}=\omega_{\text{i,c}}$, the Hamiltonian simplifies to
\bal
\hat{H}=\hat{H}_0+\hat{H}_1,
\label{Hamitonianallro2a}
\eal
where
\[\hat{H}_0/\hbar=\sum_{j,n}\omega_{n}\hat{a}^{\dagger}_j(\omega_{n})\hat{a}_j(\omega_{n})
+i\sqrt{\gamma}\bigl(\hat{a}^{\dagger}_{\text{i}}\hat{b}-\hat{b}^{\dagger}\hat{a}_{\text{i}}\bigr),
\]
\[\hat{H}_1/\hbar=-i\eta\bigl[\hat{a}_{\text{s}}\hat{b}^{\dagger}\beta(t)-\hat{a}_{\text{s}}^{\dagger}\hat{b}\beta^{*}(t)\bigr].
\]
\textit{Remark.}
The spatial integrals in the nonlinear-interaction Hamiltonian have already been
carried out. In $\hat{H}_{1}$, the phase-matching factor
$C=\int_{0}^{L} e^{i\Delta k z}\,dz$
is treated as a constant. This approximation is justified because the cavity
linewidth and pump bandwidth are chosen such that only frequency components
satisfying $\Delta k L \ll 1$ experience appreciable gain, while off–phase-matched
components are strongly suppressed. Although SFG typically supports nonlinear
interaction bandwidths of 1--10~THz~\cite{ou2007multi}, imposing this condition
(e.g., $\Delta k L \sim 0.01$) reduces the usable signal bandwidth to
0.1--1~THz~\cite{Chen:25}.

From Eq.~\eqref{Hamitonianallro2a}, the Heisenberg equations of motion for 
$\hat{a}_{j}(\omega_{n})$ and $\hat{b}$ are
\begin{align}
\dot{\hat{a}}_{\text{s}}(\omega_n,t)
&=-i\omega_{n}\hat{a}_{\text{s}}(\omega_{n},t)
+\frac{\eta}{\sqrt{T}}\hat{b}(t)\beta^{*}(t),
\label{aseqh}
\end{align}
\begin{align}
\dot{\hat{a}}_{\text{i}}(\omega_{n},t)
&=-i\omega_{n}\hat{a}_{\text{i}}(\omega_{n},t)
+\sqrt{\frac{\gamma}{T}}\hat{b}(t),
\label{ai_solh}
\end{align}
\begin{align}
\dot{\hat{b}}(t)
&=-\eta\,\hat{a}_{\text{s}}(t)\beta(t)
-\sqrt{\gamma}\,\hat{a}_{\text{i}}(t).
\label{ceq1h}
\end{align}
Solving Eqs.~\eqref{aseqh}–\eqref{ai_solh} following the method of Ref.~\cite{PhysRevA.31.3761} gives
\begin{align}
\hat{a}_{\text{s}}(\omega_{n},t)
&=e^{-i\omega_{n}(t+T/2)}
\hat{a}_{\text{s}}(\omega_{n},-T/2)
\nonumber\\&+\frac{\eta}{\sqrt{T}}
\int_{-T/2}^{t} e^{-i\omega_{n}(t-t')}
\hat{b}(t')\beta^{*}(t')\,dt',
\end{align}
\begin{align}
\hat{a}_{\text{i}}(\omega_{n},t)
&=e^{-i\omega_{n}(t+T/2)}
\hat{a}_{\text{i}}(\omega_{n},-T/2)
\nonumber\\&+\sqrt{\frac{\gamma}{T}}
\int_{-T/2}^{t} e^{-i\omega_{n}(t-t')}
\hat{b}(t')\,dt'.
\end{align}
Thus,
\begin{align}
\hat{a}_{\text{s}}(t)
&=\frac{1}{\sqrt{T}}\sum_{n}\hat{a}_{\text{s}}(\omega_{n},t)
=\hat{a}_{\text{s}}^{\text{in}}(t)
+\frac{\eta}{2}\hat{b}(t)\beta^{*}(t),
\label{amode}
\end{align}
\begin{align}
\hat{a}_{\text{i}}(t)
&=\frac{1}{\sqrt{T}}\sum_{n}\hat{a}_{\text{i}}(\omega_{n},t)
=\hat{a}_{\text{i}}^{\text{in}}(t)
+\frac{\sqrt{\gamma}}{2}\hat{b}(t).
\label{bmode}
\end{align}

Substituting Eqs.~\eqref{amode}–\eqref{bmode} into Eq.~\eqref{ceq1h} yields the Langevin equation for the cavity mode:
\begin{align}
\dot{\hat{b}}(t)
&=-\frac{\gamma}{2}\hat{b}(t)
-\eta\,\hat{a}_{\text{s}}^{\text{in}}(t)\beta(t)
-\frac{\eta^{2}}{2}\hat{b}(t)|\beta(t)|^{2}
-\sqrt{\gamma}\,\hat{a}_{\text{i}}^{\text{in}}(t),
\label{ceq2a}
\end{align}
subject to the periodic boundary condition
$\hat{b}(T/2)=\hat{b}(-T/2)$. The solution is given in 
\begin{align}
	\hat{b}(t)
	=\sum_{j}\int_{-T/2}^{T/2} g'_j(t,t')\,
	\hat{a}_{j}^{\text{in}}(t')\,dt',
	\label{boutgcapp}
\end{align}
where the kernels are
\begin{align}
	g'_j(t,t')
	&=-\left[
	\frac{e^{-\gamma T/2-\eta^2/2}}
	{1-e^{-\gamma T/2-\eta^2/2}}
	+\Theta(t-t')
	\right]
	h_j(t')\nonumber\\&\times
	\exp\!\left[-\int_{t'}^{t}
	\left(\frac{\gamma}{2}+\frac{\eta^{2}}{2}|\beta(t'')|^{2}\right)
	dt''\right],
	\label{g1aapp}
\end{align}
with
$h_{\text{s}}(t)=\eta\beta(t)$, $h_{\text{i}}(t)=\sqrt{\gamma}$
and \(\Theta\) the Heaviside step function. Using the cavity input–output relation
\be
\hat{a}_{\rm i}^{\rm out}(t)-\hat{a}_{\rm i}^{\rm in}(t)=\sqrt{\gamma}\hat{b}(t),
\label{inoutcaaapp}
\ee
together with Eq. (\ref{boutgcapp}), the idler output operator can be written directly as
\be
\hat{a}_{\rm i}^{\rm out}(t)=\sum_{j}\int_{-T/2}^{T/2} g_j(t,t')\,
\hat{a}_{j}^{\text{in}}(t')\,dt',
\label{inoutgapp}
\ee
where $g_{\rm s}(t,t')=\sqrt{\gamma} g'_{\rm s}(t,t')$ and $g_{\rm i}(t,t')=\sqrt{\gamma}g'_{\rm i}(t,t')+\delta(t,t')$.

To access the operating regime of interest, we consider the limit $\eta\sim\sqrt{\gamma T}\rightarrow 0$  ($\gamma/\Delta\omega\rightarrow 0$). In this regime, Eq. (\ref{g1aapp}) reduces to
\be
g'_j(t,t')=-2 h_j(t')/(\gamma T+\eta^2).
\ee
Consequently, the intracavity field becomes
\begin{align}
	\hat{b}(t)
	&=-\int_{-T/2}^{T/2}
	\frac{2}{\gamma T+\eta^2}
	\bigl[\eta\beta(t')\hat{a}^{\text{in}}_{\text{s}}(t')
	+\sqrt{\gamma}\,\hat{a}^{\text{in}}_{\text{i}}(t')\bigr] dt'
	\nonumber\\[3pt]
	&=-\frac{2}{\gamma T+\eta^2}
	\bigl[\eta\,\hat{A}_{\text{s}}^{\text{in}}(0)
	+\sqrt{\gamma T}\,\hat{a}^{\text{in}}_{\text{i}}(\omega_0)\bigr],
	\label{ctsolapp}
\end{align}
where the TM $\hat{A}^{\rm in}_{\rm s}(0)=\sum_{m}\beta(\omega_{-m})\hat{a}^{\rm in}_{\rm s}(\omega_m)$. In the frequency domain, 
\begin{align}
	\hat{b}(\omega_{n})
	=
	\begin{cases}
		-\dfrac{2}{\gamma T+\eta^2}
		\bigl[\eta\sqrt{T}\,\hat{A}_{\text{s}}^{\text{in}}(0)
		+\sqrt{\gamma}T\,\hat{a}^{\text{in}}_{\text{i}}(\omega_0)\bigr],
		& n=0,\\[6pt]
		0, & n\neq 0.
	\end{cases}
	\label{cwsol}
\end{align}
The idler output field follows as
\begin{align}
	\hat{a}_{\text{i}}^{\text{out}}(\omega_{n})
	=
	\begin{cases}
		\mu'_0\hat{A}_{\text{s}}^{\text{in}}(0)
		+\nu'_0\hat{a}_{\text{i}}^{\text{in}}(\omega_0),
		& n=0,\\[6pt]
		\hat{a}_{\text{i}}^{\text{in}}(\omega_{n}),
		& n\neq 0.
	\end{cases}
	\label{solfmapp}
\end{align} where \[\mu'_0=-\frac{2\eta\sqrt{\gamma T}}{\gamma T+\eta^2},\quad\nu'_0=\frac{\eta^2-\gamma T}{\gamma T+\eta^2}.\]
In particular, when $\eta=\sqrt{\gamma T}$, the output reduces to
\begin{align}
	\hat{a}_{\text{i}}^{\text{out}}(\omega_{n})
	=
	\begin{cases}
		-\hat{A}_{\text{s}}^{\text{in}}(0), & n=0,\\[3pt]
		\hat{a}_{\text{i}}^{\text{in}}(\omega_{n}), & n\neq 0.
	\end{cases}
	\label{fullcaapp}
\end{align}
\begin{figure}[t]
	\centering
	\includegraphics[width=\linewidth]{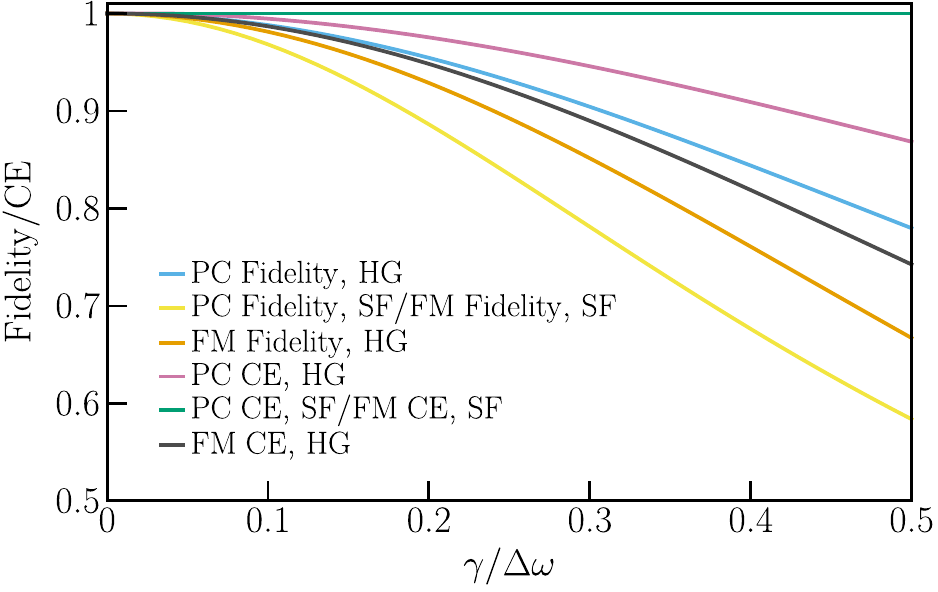}
	\caption{Comparison of the PC and FM fidelities and corresponding CEs of a
	$1\times N$ gate for second-order HG and SF pump spectra, computed using the
	same parameters as in Fig.~\ref{fig:2} of the main text.}

	\label{s2}
\end{figure}

\subsection{PC fidelity and conversion efficiency}
\label{1nfice}
We evaluate the measurement-dependent fidelity and CE of the $1\times N$
gate in the regime of small but nonzero $\gamma/\Delta\omega$, assuming
that the idler outputs are measured using a single detector, such as a
photon counter with limited spectral resolution. In this case, the measurement corresponds to
\bal
\hat{n}
&= \int_{-T/2}^{T/2} dt\;
\hat{a}_{\mathrm{i}}^{\mathrm{out}\,\dagger}(t)\,
\hat{a}_{\mathrm{i}}^{\mathrm{out}}(t)
\\&= \int_{-T/2}^{T/2} du\,du'\;
\hat{a}_{\mathrm{s}}^{\mathrm{in}\,\dagger}(u)\,
\mathcal{U}(u,u')\,
\hat{a}_{\mathrm{s}}^{\mathrm{in}}(u'),
\eal
where the associated kernel is
\be
\calU(u,u')=\!\int_{-T/2}^{T/2} dt\, g_{\mathrm{s}}^{*}(t,u)\, g_{\mathrm{s}}(t,u').
\label{calu}
\ee 
Although the idler input modes are formally included in Eq.~(\ref{inoutgapp}), we retain only the converted-mode contribution and omit the trivial vacuum terms here.

Since $\mathcal{U}$ is Hermitian and positive semidefinite by construction, we may regard 
\(
\rho = \mathcal{U} / ({\rm Tr}\,\mathcal{U})
\)
as an effective density matrix. For the ideal transformation in the time domain, $g^{\rm ideal}_{\mathrm{s}}(t,u)\propto\beta(u)$, the corresponding kernel is
\(
\mathcal{U}^{\rm ideal}(u,u') = \beta^{*}(u)\beta(u')
\),
whose normalized form
\(
\sigma = \mathcal{U}^{\rm ideal}/({\rm Tr}\,\mathcal{U}^{\rm ideal})
\)
represents a pure state.  
Borrowing the standard quantum-state fidelity, we define the
PC fidelity as
\bal
\mathcal{F}_i^{\rm PC} &= {\rm Tr}(\rho\sigma)
\\&= \frac{{\rm Tr}(\mathcal{U}\,\mathcal{U}^{\rm ideal})}{({\rm Tr}\,\mathcal{U})({\rm Tr}\,\mathcal{U}^{\rm ideal})}\\&=
\frac{
	\displaystyle 
	\int_{-T/2}^{T/2} dt \,
	\bigg|
	\int_{-T/2}^{T/2} du \; g_{\rm s}(t,u)\,\beta^{*}(u)
	\bigg|^{2}
}{
	\displaystyle 
	\iint_{-T/2}^{T/2} dt\,du \; |g_{\rm s}(t,u)|^{2}
},
\label{pcfide1n}
\eal
where we have used $ \int_{-T/2}^{T/2} du \; |\beta(u)|^{2}=1$. Correspondingly, the CE—the probability that the target mode $\hat{A}^{\mathrm{in}}_{\mathrm{s}}(0)$ is successfully converted—is
\bal
\mathcal{C}_e^{\rm PC}
&= \frac{{\rm Tr}(\mathcal{U}\,\mathcal{U}^{\rm ideal})}{({\rm Tr}\,\mathcal{U}^{\rm ideal})^2}\\&=
\int_{-T/2}^{T/2} dt \,
\bigg|
\int_{-T/2}^{T/2} du \; g_{\rm s}(t,u)\,\beta^{*}(u)
\bigg|^{2}.
\label{pcci1n}
\eal
Comparing Eqs.~(\ref{fidem1}–\ref{cem1}) with
Eqs.~(\ref{pcfide1n}–\ref{pcci1n}), we find that
$\mathcal{F}_i^{\mathrm{FM}}\leq\mathcal{F}_i^{\mathrm{PC}}$ and
$\mathcal{C}_e^{\mathrm{FM}}\leq\mathcal{C}_e^{\mathrm{PC}}$, as
guaranteed by the Cauchy–Schwarz inequality. This behavior is illustrated
in Fig.~\ref{s2}, which compares the PC and FM
fidelities and corresponding CEs for a $1\times N$ gate driven by
second-order HG and SF pump spectra, using the same parameters as in the
main text.
For the SF pump, the close agreement between the PC and FM fidelities and
CEs arises when the dominant Schmidt component of $g_{\rm s}(t,u)$—the
leading rank-1 term $\phi_0(t)\psi_0(u)$—satisfies
$\phi_0(t)\approx \mathrm{const}$ and $\psi_0(u)\approx \beta(u)$.
The SF-mode fidelity is lower than the HG-mode fidelity: HG modes vary smoothly in frequency, so adjacent bins have nearly identical amplitudes, allowing high fidelity even when the gate does not sharply resolve individual pump-frequency bins. In contrast, the SF-mode fidelity more stringently reflects the limit of the gate. 

Finally, because HD already performs a $1\times N$ mode-selection operation, applying it after another $1\times N$ gate is redundant, and we therefore omit HD-based fidelity analysis here.

\subsection{Control-mode spectral sensitivity}
\label{cmss}
To complement the fidelity analysis, we investigate the gate’s ability to distinguish the outputs produced by pump fields in two adjacent SF modes, assessed via PC. We encode the pump field in two TMs, 
$\beta(t) = (1/\sqrt{T}) e^{-i\omega_l t}$ and $(1/\sqrt{T}) e^{-i\omega_{l+1} t}$, 
corresponding to two nearby SF modes at $\omega_l$ and $\omega_{l+1}$. Since $|\beta(t)|^2=1/T$ for both cases, solving Eq.~(\ref{ceq2a}) in the frequency domain yields
\begin{equation}
	\hat{b}(\omega_{n})
	= \frac{\eta/\sqrt{T}}{i\omega_{n} - \frac{\gamma}{2} - \frac{\eta^{2}}{2T}}
	\,\hat{A}_{\rm s}^{\rm in}(n)
	+ \frac{\sqrt{\gamma}}{i\omega_{n} - \frac{\gamma}{2} - \frac{\eta^{2}}{2T}}
	\,\hat{a}_{\rm i}^{\rm in}(\omega_{n}),
	\label{cn}
\end{equation}
where $\hat{A}_{\rm s}^{\rm in}(n)
= \hat{a}_{\rm s}^{\rm in}(\omega_{n-l})$ and $\hat{a}_{\rm s}^{\rm in}(\omega_{n-l-1})$ for the two pump configurations, respectively.
Using the cavity input–output relation
\begin{equation}
	\hat{a}_{\rm i}^{\rm out}(t) - \hat{a}_{\rm i}^{\rm in}(t)
	= \sqrt{\gamma}\,\hat{b}(t),
	\label{inoutcaa}
\end{equation}
we obtain
\begin{equation}
	\hat{a}_{\rm i}^{\rm out}(\omega_{n})
	= \mu'_{n}\,\hat{A}_{\rm s}^{\rm in}(n)
	+ \nu'_{n}\,\hat{a}_{\rm i}^{\rm in}(\omega_{n}),
	\label{bouta}
\end{equation}
where
\begin{equation}
	\mu'_{n}
	= \frac{\eta\sqrt{\gamma/T}}
	{i\omega_{n} - \frac{\gamma}{2} - \frac{\eta^{2}}{2T}},
	\quad
	\nu'_{n}
	= \frac{i\omega_{n} + \frac{\gamma}{2} - \frac{\eta^{2}}{2T}}
	{i\omega_{n} - \frac{\gamma}{2} - \frac{\eta^{2}}{2T}}.
\end{equation}
Note: Under $\eta=\sqrt{\gamma T}$, the target-mode CE, given by $|\mu'_0|^2$, is unity for any pump with flat intensity $|\beta(t)|^2 = 1/T$ (e.g., all SF pumps).

\begin{figure}[t]
    \centering
 \includegraphics[width=0.6\linewidth]{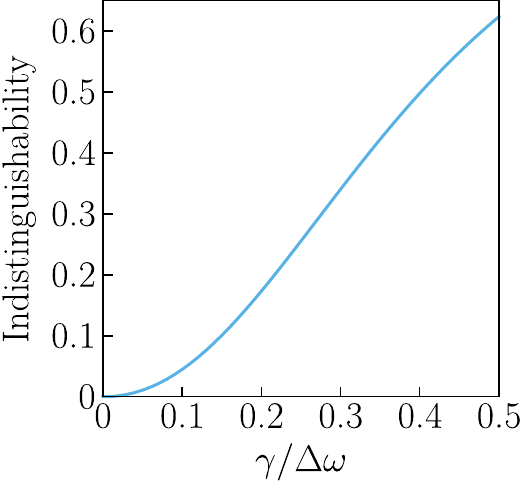}
\caption{Indistinguishability of $1\times N$ gate outputs for pumps in adjacent 
SF modes versus $\gamma/\Delta\omega$, evaluated under the condition $\eta=\sqrt{\gamma T}$.}
    \label{s3}
\end{figure}
For two pump configurations that differ by one frequency-bin index, the
corresponding frequency-domain transfer functions take the form
\(
\tilde{g}_{\rm s}^{(q)}(n,m)=\mu'_n\,\delta_{m,n-l-q+1},
\)
where $q\in\{1,2\}$ labels the two cases.
As a result, the time-domain kernel $\calU^{(q)}(u,u')$ in
Eq.~(\ref{calu}) is diagonal in the frequency-bin basis.
In the frequency representation, this implies that
$\tilde{\calU}^{(q)}$ is a diagonal matrix with elements
$(\tilde{\calU}^{(q)})_{mm}=|\mu'_{m+l+q-1}|^{2}$, since
\begin{align}
	\tilde{\calU}^{(q)}(m,m')
	&= \sum_n \tilde{g}_{\rm s}^{(q)*}(n,m)\tilde{g}_{\rm s}^{(q)}(n,m') \nonumber\\
	&= \sum_n |\mu'_n|^2\,
	\delta_{m,n-l-q+1}\delta_{m',n-l-q+1} \nonumber\\
	&= |\mu'_{m+l+q-1}|^2\,\delta_{m,m'} .
\end{align}
Each $\tilde{\calU}^{(q)}$ is Hermitian and positive semidefinite. After normalization by its trace, 
\( \rho^{(q)} = \tilde{\calU}^{(q)}/[\mathrm{Tr}\,\tilde{\calU}^{(q)}]\),
it represents a valid mixed state.
The indistinguishability between these two mixed states is quantified using the quantum
fidelity for mixed states, also known as the Uhlmann fidelity, which reduces to the classical Bhattacharyya form when
the two density matrices are diagonal and therefore commute:
\begin{equation}
\calF_i(\rho^{(1)},\rho^{(2)})
=\Big(\sum_{m}\sqrt{\rho^{(1)}(m,m)\,\rho^{(2)}(m,m)}\Big)^{2}.
\end{equation}
Substituting the explicit diagonal elements yields the indistinguishability (fidelity) between the two pump configurations:
\be
\calF_i
=\left(
     \frac{\sum_{k}|\mu'_{k}|\,|\mu'_{k+1}|}
          {\sum_{k}|\mu'_{k}|^{2}}
   \right)^{2},
\ee
under the natural assumption $\sum_{k}|\mu'_{k}|^{2} = \sum_{k}|\mu'_{k+1}|^{2}$.
The dependence of the indistinguishability on $\gamma/\Delta\omega$ is shown in 
Fig.~\ref{s3}. A high indistinguishability (low sensitivity) indicates that 
adjacent pump–frequency components produce similar outputs, which directly 
limits the fidelity achievable for an SF target mode. The fidelity results and 
the control-mode sensitivity are therefore consistent, together characterizing 
the gate’s spectral sharpness (its ability to resolve nearby pump-frequency 
components) and its practical mode selectivity.

\section{Extension to the $M\times N$ gate}
\label{mnda}
The $M\times N$ gate, based on the CSFG process, is governed by the Hamiltonian
\bal
\hat{H}/\hbar
&=\sum_m \omega_{\mathrm{i},\mathrm{c},m}\hat{b}_m^{\dagger}\hat{b}_m
+\sum_{n} (\omega_{n}+\omega_{\mathrm{s},\mathrm{c}})
\hat{a}_{\mathrm{s}}^{\dagger}(\omega_{n})
\hat{a}_{\mathrm{s}}(\omega_{n})
\\
&\quad
+\sum_{m,n}(\omega_{n}+\omega_{\mathrm{i},\mathrm{c},m})
\hat{a}^{\dagger}_{\mathrm{i},m}(\omega_{n})
\hat{a}_{\mathrm{i},m}(\omega_{n})
\\
&\quad
+i\sqrt{\gamma}\Bigl(
\sum_m\hat{a}_{\mathrm{i},m}^{\dagger}\sum_{m'}\hat{b}_{m'}
-\sum_{m'}\hat{b}_{m'}^{\dagger}\sum_m\hat{a}_{\mathrm{i},m}
\Bigr)
\\
&\quad
-i\eta\Bigl[
\hat{a}_{\mathrm{s}}
\sum_m \hat{b}_m^{\dagger}\sum_{m'}\beta_{m'}(t)
e^{-i\omega_{\mathrm{p},\mathrm{c},m'} t}
\\&\qquad-\hat{a}_{\mathrm{s}}^{\dagger}
\sum_m\hat{b}_m\sum_{m'}\beta_{m'}^{*}(t)
e^{i\omega_{\mathrm{p},\mathrm{c},m'} t}
\Bigr].
\label{Hamitonianall}
\eal
Here $\omega_{\mathrm{i},\mathrm{c},m}
=\omega_{\mathrm{i},\mathrm{c}}+m\Omega_{\mathrm{FSR}}$ and
$\omega_{\mathrm{p},\mathrm{c},m}
=\omega_{\mathrm{p},\mathrm{c}}+m\Omega_{\mathrm{FSR}}$.
The operator
\(
\hat{a}_{\mathrm{i},m}
=
\sqrt{1/T}\sum_{n}\hat{a}_{\mathrm{i},m}(\omega_{n})
\)
denotes the external idler mode, restricted to one free spectral range
around $\omega_{\mathrm{i},\mathrm{c},m}$ and coupled to the intracavity
mode $\hat{b}_m$.

In the rotating frame defined by
$\hat{a}_{\mathrm{s}}(\omega_{n})\rightarrow
\hat{a}_{\mathrm{s}}(\omega_{n})e^{-i\omega_{\mathrm{s},\mathrm{c}} t}$,
$\hat{a}_{\mathrm{i},m}(\omega_{n})\rightarrow
\hat{a}_{\mathrm{i},m}(\omega_{n})e^{-i\omega_{\mathrm{i},\mathrm{c},m} t}$,
and $\hat{b}_m\rightarrow\hat{b}_m e^{-i\omega_{\mathrm{i},\mathrm{c},m} t}$,
the Langevin equations for the cavity modes become~\cite{PhysRevA.43.543}.
\bal
\dot{\hat{b}}_k(t)
&=-\frac{\gamma}{2}\hat{b}_k(t)
-\eta\,\hat{a}_{\mathrm{s}}^{\mathrm{in}}(t)\beta_k(t)
\\&\qquad-\frac{\eta^2}{2}\sum_m \hat{b}_m(t)\beta_m^{*}(t)\beta_k(t)
-\sqrt{\gamma}\hat{a}_{\mathrm{i},k}^{\mathrm{in}}(t),
\label{ceqmulapp}
\eal
where rapidly rotating terms proportional to $e^{-im\Omega_{\mathrm{FSR}}t}$ ($m\neq0$) have been neglected.

Solving Eq.~(\ref{ceqmulapp}) in the limit $\eta\sim\sqrt{\gamma T}\rightarrow 0$  
($\gamma/\Delta\omega\rightarrow 0$) yields the analog of Eq.~(\ref{ctsolapp}):
\bal
\hat{b}_k(t)
=-\frac{2}{\gamma T+\eta^2}
\bigl[\eta\,\hat{A}_{\mathrm{s},k}^{\mathrm{in}}(0)
+\sqrt{\gamma T}\hat{a}_{\mathrm{i},k}^{\mathrm{in}}(\omega_0)\bigr]
+K_k,
\label{ctsolmapp}
\eal
where  
\[\hat{A}_{\mathrm{s},k}^{\mathrm{in}}(0)=\int_{-T/2}^{T/2} dt'\,\beta_k(t')\hat{a}_{\mathrm{s}}^{\mathrm{in}}(t')
=\sum_{l}\beta_k(\omega_{-l})\hat{a}_{\mathrm{s}}^{\mathrm{in}}(\omega_{l}),\]   
\[K_k\propto\int_{-T/2}^{T/2}dt'\sum_{m\neq k}\hat{b}_m(t')\beta_m^{*}(t')\beta_k(t').\]
Because the right-hand side of Eq.~(\ref{ctsolmapp}) is time independent, only the zero-frequency component survives:
\(
\hat{b}_m(\omega_{n})=0\) for \(n\neq 0
\),
as in Eq.~(\ref{cwsol}).  
Thus $\hat{b}_m(t)=(1/\sqrt{T})\sum_{n}\hat{b}_{m}(\omega_{n})\exp(-i\omega_{n}t)=(1/\sqrt{T})\hat{b}_m(\omega_0)$, and  
\bal
K_k&\propto\int_{-T/2}^{T/2}dt'\sum_{m\neq k}\hat{b}_m(\omega_0)
\beta_m^{*}(t')\beta_k(t')
\\&=\sum_{m\neq k}\hat{b}_m(\omega_0)\delta_{mk}
\\&=0,\nonumber
\eal
where we have used the orthogonality of the set $\{\beta_m(t)\}$.

Using the cavity input–output relations
\be
\hat{a}_{{\rm i},k}^{\rm out}(t)-\hat{a}_{{\rm i},k}^{\rm in}(t)=\sqrt{\gamma}\hat{b}_k(t),
\label{inoutcaamch}
\ee 
the idler output field in the $k$th cavity-resonance channel can be written in the frequency-bin basis as
\begin{align}
	\hat{a}_{\mathrm{i},k}^{\mathrm{out}}(\omega_{n})
	=
	\begin{cases}
		\mu'_0\,\hat{A}_{\mathrm{s},k}^{\mathrm{in}}(0)
		+\nu'_0\,\hat{a}_{\mathrm{i},k}^{\mathrm{in}}(\omega_0),
		& n=0,\\[6pt]
		\hat{a}_{\mathrm{i},k}^{\mathrm{in}}(\omega_{n}),
		& n\neq 0,
	\end{cases}
\end{align}
which is the direct multi-channel generalization of Eq.~(\ref{solfmapp}).

\subsection{Measurement-dependent fidelities and conversion efficiencies of the $M\times N$ gate}
\label{mnficesec}
In this section, we evaluate the measurement-dependent fidelities and CEs of
the $M\times N$ gate for small but nonzero $\gamma/\Delta\omega$.

Define the $M$-component operator vectors
\[
\vec b(t)=
\begin{pmatrix}
\hat b_{+(M-1)/2}(t)\\
\vdots\\
\hat b_{k}(t)\\
\vdots\\
\hat b_{-(M-1)/2}(t)
\end{pmatrix},
\quad
\bigl(\vec a_{\rm i}^{\,\rm in(out)}(t)\bigr)_k
=
\hat a_{{\rm i},k}^{\,\rm in(out)}(t).
\]
With these definitions, the coupled equations of motion
[Eqs.~(\ref{ceqmulapp})] can be written in the compact form
\be
\dot{\vec b}(t)= -\mathbb{M}(t)\,\vec b(t)-\vec d(t),
\label{eq:odecompact}
\ee
subject to the boundary condition $\vec b(T/2)=\vec b(-T/2)$. Here
\[
\mathbb{M}(t)
\equiv
\frac{\gamma}{2}\,\mathbb{I}
+
\frac{\eta^{2}}{2}\,\mathbb{J}(t),
\quad
\vec d(t)
\equiv
\sqrt{\gamma}\,\vec a_{\mathrm{i}}^{\,\mathrm{in}}(t)
+
\eta\,\hat{a}_{\rm s}^{\,\rm in}(t)\,\vec\beta(t),
\]
where \(\mathbb{I}\) is the \(M\times M\) identity matrix,
\(\mathbb{J}_{km}(t)=\beta_{m}^{*}(t)\,\beta_{k}(t)\) and \(\vec\beta(t)=\{\beta_k(t)\}^{T}\).

Eq.~\eqref{eq:odecompact} is a first-order linear differential equation with
time-dependent coefficients. Its exact solution on $[-T/2,T/2]$ can be written as
\[
\vec b(t)=\int_{-T/2}^{T/2}du\;\mathbb{K}(t,u)\,\vec d(u),
\]
where
\bal
\mathbb{K}(t,u)&=
-\mathbb{P}\!\left(t,-\tfrac{T}{2}\right)
\Big[\mathbb{I}-\mathbb{P}\!\left(\tfrac{T}{2},-\tfrac{T}{2}\right)\Big]^{-1}
\mathbb{P}\!\left(\tfrac{T}{2},u\right)
\nonumber\\&\qquad\qquad\qquad-\Theta(t-u)\,\mathbb{P}(t,u).
\eal
Here the propagator is
\[
\mathbb{P}(t,t_{0})=
\mathcal{T}\exp\!\left[-\int_{t_{0}}^{t}ds\;\mathbb{M}(s)\right],
\]
with \(\mathcal{T}\) denoting time ordering.

Using the cavity input--output relations in Eq.~(\ref{inoutcaamch}),
the idler output can be written in the input--output form
\bal
\vec a_{\rm i}^{\,\rm out}(t)
&=
\int_{-T/2}^{T/2}du\;\vec G_{\rm s}(t,u)\,\hat a_{\rm s}^{\,\rm in}(u)
\\&\qquad+
\int_{-T/2}^{T/2}du\;\mathbb{G}_{\rm i}(t,u)\,\vec a_{\rm i}^{\,\rm in}(u),
\eal
where
\[
\vec G_{\rm s}(t,u)=\sqrt{\gamma}\eta\,\mathbb{K}(t,u)\,\vec\beta(u),
\]
\[
\mathbb{G}_{\rm i}(t,u)=\gamma\,\mathbb{K}(t,u)+\delta(t-u)\,\mathbb{I}.
\]
\begin{figure}[t]
	\centering
	\includegraphics[width=\linewidth]{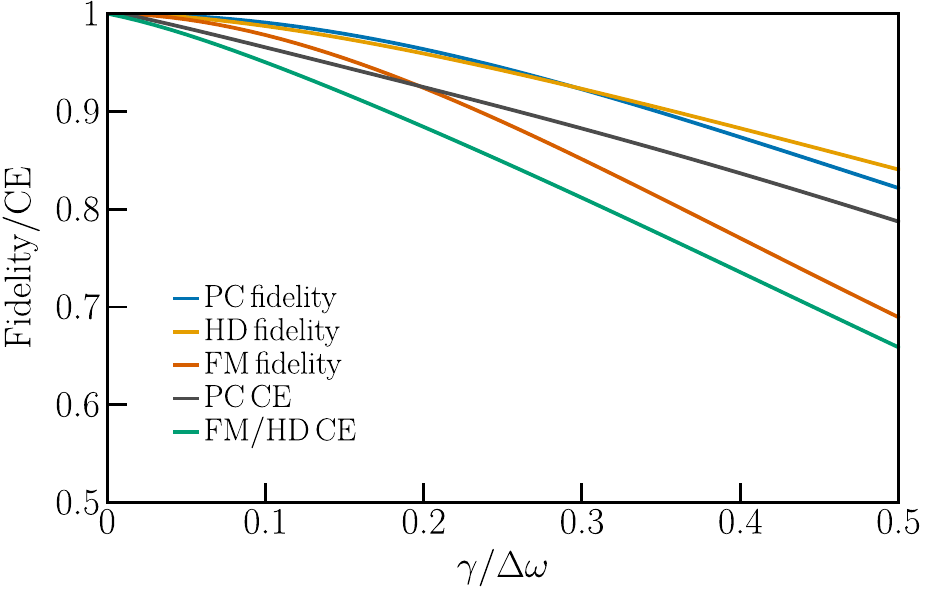}
	\caption{PC, HD, and FM fidelities and corresponding CEs of a $101\times101$ gate, evaluated with respect to the ideal identity transformation using 
		the same parameters as Fig.~\ref{fig:22} of the main text.}
	\label{s4}
\end{figure}

We assume the idler output of each channel of the $M\times N$ gate is measured by an individual photon counter after WDM. 
The single-channel PC fidelity in Eq.~(\ref{pcfide1n}) extends to the multi-channel case in a classical way as
\bal \mathcal F_i^{\mathrm{PC}} &= \frac{ \displaystyle \int_{-T/2}^{T/2} dt\; \left\| \int_{-T/2}^{T/2} du\; \bigl(\vec\beta^{\,*}(u)\circ \vec G_{\rm s}(t,u)\bigr) \right\|_{2}^{2} }{ \displaystyle \Bigl[ \iint_{-T/2}^{T/2} dt\,du\; \bigl|\vec G_{\rm s}(t,u)\bigr|^{\circ 2} \Bigr] \cdot \Bigl[ \int_{-T/2}^{T/2} du\; \bigl|\vec\beta(u)\bigr|^{\circ 2} \Bigr] } \\ &= \frac{ \displaystyle \int_{-T/2}^{T/2} dt\; \left\| \int_{-T/2}^{T/2} du\; \bigl(\vec\beta^{\,*}(u)\circ \vec G_{\rm s}(t,u)\bigr) \right\|_{2}^{2} }{ \displaystyle \iint_{-T/2}^{T/2} dt\,du\; \|\vec G_{\rm s}(t,u)\|_{2}^{2} }. 
\label{pcfidenn}
\eal
Here $\circ$ denotes the Hadamard (elementwise) product, and
$|X|^{\circ 2}$ denotes elementwise modulus squared, i.e.\ $|X|^{\circ 2}=(|X_{1}|^{2},\ldots,|X_{m}|^{2})$.
Because each channel is detected independently, PC discards all relative
phase information between channels, consistent with the invariance of the
kernel $\mathcal{U}$ in Eq.~(\ref{calu}) under channel-dependent phase shifts. The corresponding PC-based CE is given by
\be
\calC_e^{\mathrm{PC}}
=
\frac{1}{M}\int_{-T/2}^{T/2} dt\;
\left\|
\int_{-T/2}^{T/2} du\;
\bigl(\vec\beta^{\,*}(u)\circ \vec G_{\rm s}(t,u)\bigr)
\right\|_{2}^{2}
,
\label{pccenn}
\ee
where we have used $\int_{-T/2}^{T/2} du\;\|\vec\beta(u)\|_{2}^{2}=M$. 

If each output channel is instead measured by HD, with the
field projected onto the TM $1/\sqrt{T}$ (taken to be the optimal
projection for simplicity) and only the zero-frequency component retained, then
the measurement yields
\[
\vec{A}
=
\int_{-T/2}^{T/2}\frac{dt}{\sqrt{T}}\,
\vec a_{\mathrm{i}}^{\,\mathrm{out}}(t)
=
\int_{-T/2}^{T/2} du\;\vec{\mathcal H}(u)\,
\hat a_{\mathrm{s}}^{\,\mathrm{in}}(u),
\]
where
\(
\vec{\mathcal H}(u)=\int_{-T/2}^{T/2}(dt/\sqrt{T})\,\vec G_{\rm s}(t,u)
\),
and we omit the trivial idler--input vacuum term. The ideal transformation
corresponds to \(\vec{\mathcal H}^{\,\rm ideal}(u)=\vec\beta(u)\).
The associated HD fidelity is defined as the Hilbert--Schmidt overlap between
the two kernels,
\bal
\mathcal{F}_i^{\mathrm{HD}}
&=
\frac{\big|\mathrm{Tr}(\vec{\mathcal H}^{\dagger}\vec{\mathcal H}^{\,\rm ideal})\big|^{2}}
{\mathrm{Tr}(\vec{\mathcal H}^{\dagger}\vec{\mathcal H})\,
	\mathrm{Tr}(\vec{\mathcal H}^{\,\rm ideal\dagger}\vec{\mathcal H}^{\,\rm ideal}) }
\\&=
\frac{\displaystyle 
	\left|
	\int_{-T/2}^{T/2} dt
	\int_{-T/2}^{T/2} du\;
	\vec\beta^{\,\dagger}(u)\,\vec G_{\rm s}(t,u)
	\right|^{2}}
{\displaystyle 
	M\int_{-T/2}^{T/2} du\;
	\bigl\|\!\int_{-T/2}^{T/2} dt\,\vec G_{\rm s}(t,u)\bigr\|_{2}^{2}
}.
\label{hdfinn}
\eal
The HD-based CE coincides with the FM-based CE, since
$\big|\mathrm{Tr}(\vec{\mathcal H}^{\dagger}\vec{\mathcal H}^{\,\rm ideal})\big|^{2}
=
\big|\mathrm{Tr}(\tilde{G}_{\rm s}^{\dagger}\tilde{G}_{\rm s}^{\,\rm ideal})\big|^{2}$,
as follows from comparing Eqs.~(\ref{mnfmfide}) and (\ref{hdfinn}). Moreover,
comparing Eqs.~(\ref{mnfmfide})–(\ref{fmcenn}) with
Eqs.~(\ref{pcfidenn})–(\ref{hdfinn}), we find that the FM-based fidelity and CE
are always less than or equal to their measurement-dependent counterparts, as
ensured by the Cauchy--Schwarz inequality.
This trend is illustrated in Fig.~\ref{s4}, which compares the FM-based fidelity
and CE with their PC- and HD-based counterparts for a $101\times101$ gate,
evaluated using the same parameters as in the main-text calculation, confirming
the FM figure of merit as a conservative benchmark for gate performance.




\bibliography{apssamp.bib}


\end{document}